\documentclass[epj]{svjour}
\usepackage{pstcol}
\usepackage{feynmf}
\usepackage{amsmath,amssymb,amsfonts}
\begin{document}
\title{Renormalization of QED with planar binary trees}
\author{Christian Brouder\inst{1}, Alessandra Frabetti\inst{2}}
\institute{Laboratoire de Min\'eralogie-Cristallographie, CNRS UMR7590,
 Universit\'es Paris 6 et 7, IPGP, 4 place Jussieu, 
  75252 Paris Cedex 05,
  France. {\small \tt brouder@lmcp.jussieu.fr} \and
Institut de Recherche Math\'ematique Avanc\'ee,
CNRS UMR 7501, Universit\'e Louis Pasteur, 7 rue Ren\'e Descartes,
67084 Strasbourg Cedex, France.
{\small \tt frabetti@math.u-strasbg.fr}}
\date{\today}
\abstract{The renormalized photon and electron propagators
are expanded over planar binary trees.
Explicit recurrence solutions are given for the terms
of these expansions.
In the case of massless Quantum
Electrodynamics (QED), 
the relation between renormalized and bare expansions
is given in terms of a Hopf algebra structure.
For massive quenched QED, the relation between renormalized
and bare expansions is given explicitly.
} 
\maketitle
\newcommand{\dd}{\mathrm{d}}
\newcommand{\Id}{\mathrm{Id}}
\newcommand{\varpsi}{\psi}
\newcommand{\Sstar}{S_\star}
\newcommand{\discr}{{\cal{D}}}
\newcommand{\DeltaP}{\Delta^P}
\newcommand{\nablaP}{\nabla}
\newcommand{\veeb}{
\setlength{\unitlength}{2pt}
\psset{unit=2pt}
\psset{runit=2pt}
\psset{linewidth=0.15}
\begin{pspicture}(0,0)(4,3)
\psline(1.5,0)(0,3)
\psline(1.5,0)(3,3)
\psdots[dotstyle=o](1.5,0)
\end{pspicture}}
\newcommand{\veen}{
\setlength{\unitlength}{2pt}
\psset{unit=2pt}
\psset{runit=2pt}
\psset{linewidth=0.15}
\begin{pspicture}(0,0)(4,3)
\psline(1.5,0)(0,3)
\psline(1.5,0)(3,3)
\psdots[dotstyle=*](1.5,0)
\end{pspicture}}

\renewcommand{\|}{\setlength{\unitlength}{2pt}
\psset{unit=2pt}
\psset{runit=2pt}
\psset{linewidth=0.2}
\begin{pspicture}(0,0)(2,2)
\psline(1,0)(1,2)
\end{pspicture}}

\newcommand{\Y}{
\setlength{\unitlength}{2pt}
\psset{unit=2pt}
\psset{runit=2pt}
\psset{linewidth=0.2}
\begin{pspicture}(0,0)(2,3)
\psline(1,0)(1,2)
\psline(1,2)(0,3)
\psline(1,2)(2,3)
\end{pspicture}}

\newcommand{\tunb}{\setlength{\unitlength}{2pt}
\psset{unit=2pt}
\psset{runit=2pt}
\psset{linewidth=0.2}
\begin{pspicture}(0,0)(2,2)
\psdots[dotstyle=o](1,1)
\end{pspicture}}

\newcommand{\tunn}{\setlength{\unitlength}{2pt}
\psset{unit=2pt}
\psset{runit=2pt}
\psset{linewidth=0.2}
\begin{pspicture}(0,0)(2,2)
\psdots[dotstyle=*](1,1)
\end{pspicture}}

\newcommand\deuxun{
\setlength{\unitlength}{2pt}
\psset{unit=2pt}
\psset{runit=2pt}
\psset{linewidth=0.2}
\begin{pspicture}(0,0)(5,5)
\psline(3,0)(3,2)
\psline(3,2)(1,4)
\psline(3,2)(4,3)
\psline(2,3)(3,4)
\end{pspicture}}

\newcommand{\tdeuxunn}{\setlength{\unitlength}{3pt}
\psset{unit=3pt}
\psset{runit=3pt}
\psset{linewidth=0.2}
\begin{pspicture}(0,0)(2,2)
\psline(1,0)(1,2)
\psdots[dotstyle=*](1,0)
\psdots[dotstyle=o](1,2)
\end{pspicture}}

\newcommand\deuxdeux{
\setlength{\unitlength}{2pt}
\psset{unit=2pt}
\psset{runit=2pt}
\psset{linewidth=0.2}
\begin{pspicture}(1.5,0)(5,5)
\psline(3,0)(3,2)
\psline(3,2)(5,4)
\psline(3,2)(2,3)
\psline(4,3)(3,4)
\end{pspicture}}

\newcommand{\tdeuxdeuxn}{\setlength{\unitlength}{3pt}
\psset{unit=3pt}
\psset{runit=3pt}
\psset{linewidth=0.2}
\begin{pspicture}(0,0)(2,2)
\psline(1,0)(1,2)
\psdots[dotstyle=*](1,0)
\psdots[dotstyle=*](1,2)
\end{pspicture}}

\newcommand{\tdeuxdeuxb}{\setlength{\unitlength}{3pt}
\psset{unit=3pt}
\psset{runit=3pt}
\psset{linewidth=0.2}
\begin{pspicture}(0,0)(2,2)
\psline(1,0)(1,2)
\psdots[dotstyle=o](1,0)
\psdots[dotstyle=*](1,2)
\end{pspicture}}

\newcommand{\ttroisquatreb}{\setlength{\unitlength}{3pt}
\psset{unit=3pt}
\psset{runit=3pt}
\psset{linewidth=0.2}
\begin{pspicture}(0,0)(2,4)
\psline(1,0)(1,2)
\psline(1,2)(1,4)
\psdots[dotstyle=o](1,0)
\psdots[dotstyle=*](1,2)
\psdots[dotstyle=o](1,4)
\end{pspicture}}

\newcommand{\ttroiscinqb}{\setlength{\unitlength}{3pt}
\psset{unit=3pt}
\psset{runit=3pt}
\psset{linewidth=0.2}
\begin{pspicture}(0,0)(2,4)
\psline(1,0)(1,2)
\psline(1,2)(1,4)
\psdots[dotstyle=o](1,0)
\psdots[dotstyle=*](1,2)
\psdots[dotstyle=*](1,4)
\end{pspicture}}

\newcommand{\ttroisunn}{\setlength{\unitlength}{3pt}
\psset{unit=3pt}
\psset{runit=3pt}
\psset{linewidth=0.2}
\begin{pspicture}(0,0)(2,2)
\psline(1,0)(0,2)
\psline(1,0)(2,2)
\psdots[dotstyle=*](1,0)
\psdots[dotstyle=o](0,2)
\psdots[dotstyle=o](2,2)
\end{pspicture}}

\newcommand{\ttroisdeuxn}{\setlength{\unitlength}{3pt}
\psset{unit=3pt}
\psset{runit=3pt}
\psset{linewidth=0.2}
\begin{pspicture}(0,0)(2,4)
\psline(1,0)(1,2)
\psline(1,2)(1,4)
\psdots[dotstyle=*](1,0)
\psdots[dotstyle=o](1,2)
\psdots[dotstyle=*](1,4)
\end{pspicture}}

\newcommand{\ttroistroisn}{\setlength{\unitlength}{3pt}
\psset{unit=3pt}
\psset{runit=3pt}
\psset{linewidth=0.2}
\begin{pspicture}(0,0)(2,2)
\psline(1,0)(0,2)
\psline(1,0)(2,2)
\psdots[dotstyle=*](1,0)
\psdots[dotstyle=o](0,2)
\psdots[dotstyle=*](2,2)
\end{pspicture}}

\newcommand{\ttroisquatren}{\setlength{\unitlength}{3pt}
\psset{unit=3pt}
\psset{runit=3pt}
\psset{linewidth=0.2}
\begin{pspicture}(0,0)(2,4)
\psline(1,0)(1,2)
\psline(1,2)(1,4)
\psdots[dotstyle=*](1,0)
\psdots[dotstyle=*](1,2)
\psdots[dotstyle=o](1,4)
\end{pspicture}}

\newcommand{\ttroiscinqn}{\setlength{\unitlength}{3pt}
\psset{unit=3pt}
\psset{runit=3pt}
\psset{linewidth=0.2}
\begin{pspicture}(0,0)(2,4)
\psline(1,0)(1,2)
\psline(1,2)(1,4)
\psdots[dotstyle=*](1,0)
\psdots[dotstyle=*](1,2)
\psdots[dotstyle=*](1,4)
\end{pspicture}}

\newcommand\troisun{
\setlength{\unitlength}{2pt}
\psset{unit=2pt}
\psset{runit=2pt}
\psset{linewidth=0.2}
\begin{pspicture}(0,0)(5,5)
\psline(3,0)(3,2)
\psline(3,2)(0,5)
\psline(3,2)(4,3)
\psline(2,3)(3,4)
\psline(1,4)(2,5)
\end{pspicture}}

\newcommand\troisdeux{
\setlength{\unitlength}{2pt}
\psset{unit=2pt}
\psset{runit=2pt}
\psset{linewidth=0.2}
\begin{pspicture}(1,0)(5,5)
\psline(3,0)(3,2)
\psline(3,2)(1,4)
\psline(3,2)(4,3)
\psline(2,3)(4,5)
\psline(3,4)(2,5)
\end{pspicture}}

\newcommand\troistrois{
\setlength{\unitlength}{2pt}
\psset{unit=2pt}
\psset{runit=2pt}
\psset{linewidth=0.2}
\begin{pspicture}(0.7,0)(5,5)
\psline(3,0)(3,2)
\psline(3,2)(0.5,4.5)
\psline(1.5,3.5)(2.5,4.5)
\psline(3,2)(5.5,4.5)
\psline(4.5,3.5)(3.5,4.5)
\end{pspicture}}

\newcommand\troisquatre{
\setlength{\unitlength}{2pt}
\psset{unit=2pt}
\psset{runit=2pt}
\psset{linewidth=0.2}
\begin{pspicture}(1.5,0)(4,5)
\psline(3,0)(3,2)
\psline(3,2)(5,4)
\psline(3,2)(2,3)
\psline(4,3)(2,5)
\psline(3,4)(4,5)
\end{pspicture}}

\newcommand\troiscinq{
\setlength{\unitlength}{2pt}
\psset{unit=2pt}
\psset{runit=2pt}
\psset{linewidth=0.2}
\begin{pspicture}(1.5,0)(5,5)
\psline(3,0)(3,2)
\psline(3,2)(6,5)
\psline(3,2)(2,3)
\psline(4,3)(3,4)
\psline(5,4)(4,5)
\end{pspicture}}


\newcommand{\rb}{
\setlength{\unitlength}{2pt}
\psset{unit=2pt}
\psset{runit=2pt}
\psset{linewidth=0.2}
\begin{pspicture}(0,0)(3,3)
\psline(1,0)(1,2)
\psdots[dotstyle=o](1,3)
\end{pspicture}}

\newcommand{\rn}{
\setlength{\unitlength}{2pt}
\psset{unit=2pt}
\psset{runit=2pt}
\psset{linewidth=0.2}
\begin{pspicture}(0,0)(3,3)
\psline(1,0)(1,2)
\psdots[dotstyle=*](1,3)
\end{pspicture}}

\newcommand{\Yb}{
\setlength{\unitlength}{2pt}
\psset{unit=2pt}
\psset{runit=2pt}
\psset{linewidth=0.2}
\begin{pspicture}(0,0)(3,4)
\psline(1.5,0)(1.5,2)
\psline(1.2,2.8)(0,4)
\psline(1.8,2.8)(3,4)
\psdots[dotstyle=o](1.5,2)
\end{pspicture}}

\newcommand{\Yn}{
\setlength{\unitlength}{2pt}
\psset{unit=2pt}
\psset{runit=2pt}
\psset{linewidth=0.2}
\begin{pspicture}(0,0)(3,4)
\psline(1.5,0)(1.5,2)
\psline(1.2,2.8)(0,4)
\psline(1.8,2.8)(3,4)
\psdots[dotstyle=*](1.5,2)
\end{pspicture}}

\newcommand\deuxunb{
\setlength{\unitlength}{2pt}
\psset{unit=2pt}
\psset{runit=2pt}
\psset{linewidth=0.2}
\begin{pspicture}(0,0)(5,5)
\psline(2.5,0)(2.5,2)
\psline(2.2,3)(0,5)
\psline(1,4)(2,5)
\psline(2.8,2.8)(4,4)
\psdots[dotstyle=o](2.5,2)
\end{pspicture}}

\newcommand\deuxdeuxb{
\setlength{\unitlength}{2pt}
\psset{unit=2pt}
\psset{runit=2pt}
\psset{linewidth=0.2}
\begin{pspicture}(1.5,0)(5,5)
\psline(2.5,0)(2.5,2)
\psline(2.2,2.8)(1,4)
\psline(2.8,2.8)(5,5)
\psline(4,4)(3,5)
\psdots[dotstyle=o](2.5,2)
\end{pspicture}}


\newcommand{\YL}{
\setlength{\unitlength}{2pt}
\psset{unit=2pt}
\psset{runit=2pt}
\psset{linewidth=0.2}
\begin{pspicture}(0,0)(4,3)
\psline(3,0)(1,2)
\psline(2,1)(3,2)
\end{pspicture}}

\newcommand\deuxunL{
\setlength{\unitlength}{2pt}
\psset{unit=2pt}
\psset{runit=2pt}
\psset{linewidth=0.2}
\begin{pspicture}(0,0)(5,4)
\psline(4,0)(1,3)
\psline(3,1)(4,2)
\psline(2,2)(3,3)
\end{pspicture}}

\newcommand\deuxdeuxL{
\setlength{\unitlength}{2pt}
\psset{unit=2pt}
\psset{runit=2pt}
\psset{linewidth=0.2}
\begin{pspicture}(0,0)(5,4)
\psline(3,0)(1,2)
\psline(2,1)(4,3)
\psline(3,2)(2,3)
\end{pspicture}}


\section{Introduction}
Wightman commented \cite{Wightman}:
``Renormalization Theory has a history of egregious errors by
distinguished savants. It has a justified reputation of 
perversity; a method that works up to 13-th order in the
perturbation series fails in the 14-th order.''
Although renormalization theory is often considered to
be well understood, it is still a difficult subject plagued with
considerable combinatorial complexity.

However, renormalization is not just a recipe to extract
a finite part from an infinite integral, since it has
a deep physical meaning. It was a guide to
elaborate the theories of weak and strong interactions.
It can be used to build consistent Lagrangians: the
minimal coupling Lagrangian of scalar electrodynamics
misses a quartic term which is reintroduced by 
renormalization \cite{Itzykson}. Renormalization
is also linked to the irreversibility of 
the macroscopic Universe \cite{Anselmi}. Other
arguments in favor of renormalization have been
given by Jackiw \cite{Jackiw}. Moreover, thanks
to the work of Kreimer and Connes 
\cite{Kreimer98,Connes,ConnesK,CKI,CKII},
renormalization has become a beautiful mathematical
theory.

There is a class of physicists who think that a
quantum field theory is entirely contained in its
Feynman diagrams. Since two members of this class,
Veltman and 't Hooft, have been recently awarded the
Nobel prize in physics, it seems rather out of place to
try to formulate quantum field theory without 
Feynman diagrams.
Nevertheless, we shall pursue our attempt to build such a 
theory by studying how renormalization can be implemented
in the framework of planar binary trees.

The typical equations of QED are `implicit' functional
derivative equations.
In \cite{BrouderEPJC2}, the tree expansion method enabled
us to establish an explicit recursive solution for the
photon and electron propagators.
For renormalized QED including mass renormalization
the functional equations change drastically, and it is not
a priori clear that a similar explicit recursive solution
can be given.
In this paper, we obtain an explicit recursive solution
for renormalized QED.
In other words, the terms of the tree expansion
for the renormalized photon and electron propagators
are written as an integral of renormalized terms for
smaller trees. In the case of quenched QED, this
recursive equation is transformed into an explicit
relation between the renormalized and bare terms.
Finally, we exhibit a Hopf algebra that 
encodes the renormalization of massless QED.
This Hopf algebra is neither commutative nor
cocommutative.

The plan of the paper is the following. We first introduce
the problem of renormalization from different points of
view, then we give the essential equations for the
renormalization of QED. The tree expansion method is
presented in detail, and used to derive the recursive
equations for the renormalized electron and photon
propagators. Then detailed relations between the 
renormalized and bare propagators of quenched QED are
given. Finally the Hopf algebra of massless QED is
described. A first appendix gives some proofs, a second
one gives the relation between renormalized and bare
propagators for massless QED. A last appendix expresses
the smallest planar binary trees as a sum of Feynman diagrams.

A planar binary tree can be considered as a sum of 
Feynman diagrams (see appendix 3). To some extent, this sum is natural.
For instance, all the diagrams obtained by permutations of
the photon external lines of a Feynman diagrams 
cannot be distinguished by an experiment. Therefore, it is 
natural to regroup all these diagrams under a single symbol.
However, when renormalization comes into play, each Feynman
diagrams is renormalized by a number of Feynman subdiagrams
multiplied by scalar counterterms.
It is not a priori clear that these subdiagrams and counterterms
can be
grouped into the same trees as the original Feynman diagrams.
The main point of this paper is to show that this is
possible for QED. In other words, we provide an algebraic
structure on planar binary trees which is
compatible with renormalization.

In this paper, we consider only the renormalization
of ultraviolet divergences, and we assume that infrared
divergences are regularized, for instance by introducing
a photon mass.

\section{Renormalization}

An enormous literature has been devoted to the
renormalization theory. The reader is deferred to
\cite{Collins,Brown,Schweber} and references therein
for historical and conceptual aspects of renormalization.
Here we shall concentrate on its technical aspects.
Renormalization theory can be considered from at least three
different points of view: the Dyson-Salam method,
the extension of distributions and the product of 
distributions. 

\subsection{The Dyson point of view}
According to the first point of view, perturbative
quantum field theory yields divergent integrals
in Fourier space,
and renormalization is a technique intended to extract
a finite part from them.
A picture of how this could be achieved 
was first given by Dyson in 1949 \cite{Dyson} and Salam
\cite{Salam1,Salam2} in 1951. Explicit formulas were
proposed by Bogoliubov and Parasiuk \cite{BP57}
and finally proved by Hepp \cite{Hepp,HeppB}.
This method is general, in the sense that it can
be used for any quantum field theory, whether renormalizable
or not.

To understand this renormalization process, it is very useful
to treat a one-dimensional toy model of overlapping divergences
proposed by Kreimer \cite{Kreimer98}.
Let
\begin{eqnarray*}
f(x,y,c)=\frac{x}{x+c}\frac{1}{x+y}\frac{y}{y+c}.
\end{eqnarray*}
We want to give a meaning to the integral
\begin{eqnarray*}
I(c)=\int_1^\infty \dd x \int_1^\infty \dd y f(x,y,c).
\end{eqnarray*}

Power counting is applied as follows. If we substitute
$\lambda x$ to $x$ and take the limit $\lambda\rightarrow \infty$,
we see that $I(c)$ varies as $\lambda^0=1$, and the integral
is logarithmically divergent for $x$. Similarly, it is
logarithmically divergent for $y$. If both $x$ and $y$
are multiplied by $\lambda$, the integral
$I(c)$ varies as $\lambda^1$ in the limit $\lambda\rightarrow \infty$.
Then, $I(c)$ is linearly divergent for the variables $x,y$.
In the Dyson-Salam renormalization scheme, we first fix $y$ in
$f(x,y,c)$, we keep the part of
$f(x,y,c)$ which does not depend on $x$ (i.e. $y/(y+c)$)
and we take the value of the rest (i.e. $x/((x+c)(x+y))$) at
$c=0$ and $y=0$ (i.e. $1/x$). The product of these two
factors (i.e. $y/(x(y+c))$) is called a counterterm and is
subracted from $f(x,y,c)$ to remove the logarithmic divergence
for $x$. This procedure produces the term
\begin{eqnarray*}
f(x,y,c)-\frac{y}{x(y+c)}=-\frac{y}{y+c}\frac{xy+xc+yc}{x(x+c)(x+y)},
\end{eqnarray*}
which is now convergent for the integral over $x$
(it varies as $\lambda^{-1}$ by power counting).
If we make the same subtraction while fixing the variable $x$ we
obtain, subtracting both counterterms:
\begin{eqnarray*}
g(x,y,c)&=&
f(x,y,c)-\frac{y}{x(y+c)}-\frac{x}{y(x+c)}\\&&\hspace*{-8mm}=-
\frac{x^2y^2+xy^3+yx^3+cy^3+cx^3+cyx^2+cxy^2}{x(x+c)(x+y)(y+c)y}.
\end{eqnarray*}
The result is disappointing, because $g(x,y,c)$ is now
linearly divergent if $x$ is multiplied by $\lambda$, if $y$
is multiplied by $\lambda$ and if $x$ and $y$ are both multiplied
by $\lambda$. In other words, $g(x,y,c)$ is still more divergent
than $f(x,y,c)$.
The miracle happens when we subtract the global linear divergence
of $g(x,y,c)$. The final term
\begin{eqnarray*}
\bar{f}(x,y,c)&=&g(x,y,c)-g(x,y,0)-c\frac{\partial g(x,y,0)}{\partial c}\\
&=&
-c^2\frac{xy+cx+cy}{x(x+c)(x+y)(y+c)y},
\end{eqnarray*}
is now absolutely convergent for $x$, for $y$ and for $x,y$.

\subsection{The extension of distributions}
From a mathematical point of view, renormalization theory
can be considered as the problem of extending a distribution
to a larger domain.

The standard example is $1/x$. If $\phi(x)$ is a test function
that vanishes at 0, then 
\begin{eqnarray*}
\int_{-\infty}^{\infty} \dd x \frac{\phi(x)}{x}
\end{eqnarray*}
exists. The question is how is it possible to extend this 
distribution to general test functions. The existence of this
extension is ensured by the Hahn-Banach theorem \cite{ReedSimonII}
and a formula for such extensions is  
\begin{eqnarray*}
\int_{|x|<a} \dd x \frac{\phi(x)-\phi(0)}{x}+
\int_{|x|>a} \dd x \frac{\phi(x)}{x},
\end{eqnarray*}
for any positive parameter $a$. Hence
various extensions are possible, that
are parametrized by $a$.
Notice that the difference between two such integrals for $a$ and $a'$
is $(\log a'- \log a) \phi(0)$. Therefore, as distributions,
two extensions of $1/x$ differ by $\log \Lambda \delta(x)$
for some $\Lambda$. The peculiarity of quantum field theory
is that $\Lambda$ can be determined by experiment.

The mathematical conditions for the existence of such an extension
were investigated by Malgrange \cite{Malgrange} and
Estrada \cite{Estrada}.

This extension method can be used to calculate,
in some cases, the product of two distributions.
For instance, by Fourier transform, it can be shown
that $\delta(x-a)\delta(x)=\delta(a)$, for $a\not=0$.
However, if $a=0$, the Fourier transform of the
product diverges. This is exactly the same type
of divergence that is met in the usual presentation
of renormalization.  The product of distributions
$\delta(x)^2$ is zero for $x\not=0$, thus a possible
extension is $\delta(x)^2=C\delta(x)$, where 
$C$ is a constant determined by experiment. 

In quantum field theory, causality, 
Poincar\'e invariance and unitarity
were used by Stuekelberg and coll. to provide
a prescription to carry out this extension
\cite{Rivier,StueckelbergR,StueckelbergG,StueckelbergP}.
Bogoliubov and coll. systematized this
construction \cite{BP57,BS55,BS56,Bogoliubov},
which took its final form with Epstein and
Glaser \cite{Epstein}. Nowadays, the extension method
is called the ``causal approach'', and the case
of QED is treated in detail in Ref.\cite{Scharf}.

A (correct) proof of the validity of Bogoliubov's method 
was finally given by Hepp \cite{Hepp} in 1966 and by Zimmermann 
\cite{Zimmermann} in 1969.

Recently, the causal approach has been reintepreted in
terms of microlocal analysis \cite{Brunetti2}.
This enabled these authors to adapt the causal approach
to quantum field theory in curved spacetime.

An up to date and clear presentation of the causal approach
can be found in Ref.\cite{Pinter}.

\subsection{The product of distributions}

The most radical approach to renormalization would
be to define a product of distributions, which
could lead to a nonlinear theory of distributions. Schwartz
has shown that this is impossible in general \cite{Schwartz},
but the notion of distribution can be extended to a more 
general kind of functions which can be multiplied. 
For a comparison with experimental results, we must
project these new functions back onto standard distributions.

This approach was investigated by various authors
\cite{Colombeau,Colombeau2,Shelkovich,Bremermann,Oberguggenberger1,%
Oberguggenberger2,Jelinek,Li,Meyer}. The main drawback of
these new generalized functions is that they lead to very intricate 
calculations. For instance, it is not difficult to show that
\cite{Mikunsinski}
\begin{eqnarray*}
{\left(\frac{1}{x}\right)}^2=
\frac{1}{x^2}+
\pi^2\delta(x)^2,
\end{eqnarray*}
but the computation of $(1/x)^3$ is already intractable.
To understand this striking identity, we start from the
continuous function $f(x)=x\log|x|-x$, which defines
a distribution by $\int dx f(x)\phi(x)$ for any test
function $\phi(x)$. Then the distribution $1/x$ is
defined as $d^2f/dx^2$, the distribution 
$1/x^2$ is $-d^3f/dx^3$, and $(1/x)^2$ is the product
of the distribution $1/x$ with itself.

In spite of their complexity, these new generalized
functions have found some applications in physics
\cite{Embacher,Clarke,Steinbauer,Vickers}.
For instance, a definite value could be given to
the curvature of a cone at its apex \cite{Clarke}.

Notice that, as for the extension of distributions,
microlocal analysis is taking a growing importance
in the study of the new generalized functions \cite{Dapic}.

\section{Renormalization of QED}

QED was renormalized to all orders by Dyson \cite{Dyson}.
We can now interpret his prescriptions in the framework
of the Schwinger equations. It is standard to define
free, bare and renormalized propagators.
The free electron Green function $S^0(q)$ is the
Green function for an electron without electromagnetic interaction.
The bare electron Green function $S(q)$ is the Green function
for an electron with electromagetic interaction, but without
renormalization. In the perturbation expansion of $S(q)$,
all terms (except the first one) are infinite.
The renormalized Green function $\bar S(q)$ is the Green
function for an electron with electromagnetic interaction,
after renormalization.
Similarly we define $D^0(q)$, $D(q)$ and $\bar D(q)$ as the
free, bare and renormalized photon Green functions.
\footnote{Strictly speaking, a propagator is a 
one-particle Green function, but in this paper propagator
and Green function will be used indifferently. For simplicity,
fermions (electrons+positrons) are called electrons.}

\subsection{The free propagators}

The free electron propagator is 
\begin{eqnarray*}
S^0(q)={(\gamma\cdot q-m+i\epsilon)}^{-1}.
\end{eqnarray*}
The scalar product is defined by 
\begin{eqnarray*}
\gamma\cdot q=\sum_{\lambda\mu} \gamma^\lambda g_{\lambda\mu}q^\mu,
\end{eqnarray*}
where the pseudo-metric tensor $g_{\lambda\mu}$ is
\begin{eqnarray*}
g=\left( \begin{array}{cccc} 1 & 0 & 0 & 0\\
                            0 &-1 & 0 & 0\\
                            0 & 0 &-1 & 0\\
                            0 & 0 & 0 &-1
                          \end{array} \right).
\end{eqnarray*}

All the electron propagators $S^0(q)$, $S(q)$ and $\bar S(q)$
are $4\times 4$ complex matrix functions of the
4-vector $q$. If $I$ is the $2\times 2$ identity matrix
and $\sigma_x$, $\sigma_y$, $\sigma_z$ the Pauli matrices
\begin{eqnarray*}
\sigma_x=\left( \begin{array}{cc} 0 & 1 \\
                                 1 & 0 \end{array} \right)
\quad
\sigma_y=\left( \begin{array}{cc} 0 &-i \\
                                 i & 0 \end{array} \right)
\quad
\sigma_z=\left( \begin{array}{cc} 1 & 0 \\
                                 0 &-1 \end{array} \right),
\end{eqnarray*}
the Dirac matrices can be written
$\gamma^0=\left( \begin{array}{cc} I & 0 \\
                                  0 &-I \end{array} \right)$
\begin{eqnarray*}
\gamma^1=\left( \begin{array}{cc} 0 & \sigma_x \\
                                 -\sigma_x & 0 \end{array} \right)
\quad
\gamma^2=\left( \begin{array}{cc} 0 & \sigma_y \\
                                 -\sigma_y & 0 \end{array} \right)
\quad
\gamma^3=\left( \begin{array}{cc} 0 & \sigma_z \\
                                 -\sigma_z & 0 \end{array} \right).
\end{eqnarray*}                                                              

The free photon propagator $D^0(q)$ is a complex $4\times 4$
matrix with components $D^0_{\mu\nu}(q)$ defined by
\begin{eqnarray*}
D^0_{\mu\nu}(q)&=&
-\frac{g_{\mu\nu}}{q^2+i\epsilon}+
   (1-1/\xi)\frac{q_\mu q_\nu}{(q^2+i\epsilon)^2}.
\end{eqnarray*}
The term $1/\xi$ was introduced by Heisenberg to make
$D^0(q)$ non singular. The Green function used in classical
electrodynamics is $D^{0T}(q)$ defined as
\begin{eqnarray*}
D^{0T}_{\mu\nu}(q)&=&
-\frac{g_{\mu\nu}}{q^2+i\epsilon}+
   \frac{q_\mu q_\nu}{(q^2+i\epsilon)^2}.
\end{eqnarray*}

A tensor $T_{\mu\nu}(q)$ such that $q^\mu T_{\mu\nu}(q)=0$
is called transverse. It can be checked that $D^{0T}_{\mu\nu}(q)$
is transverse, and non singular in the space of transverse
tensors.

Up to an eventual factor $i$, 
the expressions for the free Green functions
$S^0(q)$ and $D^0_{\mu\nu}(q)$
are standard (see, e.g. Ref.\cite{Itzykson} p.93 and p.36,
Ref.\cite{Sterman} p.184 and p.190,
Ref.\cite{Ticciati} p.218 and p.253, for a
complete description and a derivation).

\subsection{The bare propagators}

The Schwinger equations for bare
electron and photon propagators were given
by Bogoliubov and Shirkov \cite{Bogoliubov} and
transformed into the following integral equations 
in \cite{BrouderEPJC2}.

\begin{eqnarray}
S(q) &=& S^0(q)
  \nonumber\\&& \hspace*{-1mm}
 +ie_0^2 S^0(q)
  \int\frac{\dd^4p}{(2\pi)^4}
  \gamma^\lambda D_{\lambda\lambda'}(p)
  \frac{\delta S(q-p)}{e_0\delta A^0_{\lambda'}(p)},\\
D_{\mu\nu}(q) &=& D^0_{\mu\nu}(q)\label{SchbareS}
 \nonumber\\&& \hspace*{-10mm}
  -ie_0^2 D^0_{\mu\lambda}(q)
  \int\frac{\dd^4p}{(2\pi)^4} \mathrm{tr}\big[\gamma^\lambda
  \frac{\delta S(p)}{e_0\delta A^0_{\lambda'}(-q)}
   \big] D_{\lambda'\nu}(q),\label{SchbareD}
\end{eqnarray}
where $A^0_{\lambda}(p)$  is an external electromagnetic field
and $\frac{\delta S(q)}{\delta A^0_{\lambda}(p)}$
is the functional derivative evaluated at $A^0_{\lambda}(p)=0$.

The longitudinal part of the bare photon Green function
is not modified by the interaction \cite{Itzykson},
and $D_{\lambda\mu}(q)$ can be written as
the sum of its transverse and its longitudinal parts:
\begin{eqnarray}
D_{\lambda\mu}(q)&=&D^T_{\lambda\mu}(q)-
\frac{1}{\xi} \frac{q_\lambda q_\mu}{(q^2+i\epsilon)^2}\label{DT},
\end{eqnarray}
where $D^T_{\lambda\mu}(q)$ is transverse.
In Eq.(\ref{SchbareD}), the photon propagator
$D_{\lambda'\nu}(q)$ is not integrated,
it just multiplies the integral. 
This is not very 
convenient and it will be useful to introduce the bare
vacuum polarization, denoted
$\Pi_{\lambda\mu}(q)$ 
and defined by
\begin{eqnarray}
{[D^{-1}]}_{\lambda\mu}(q)
&=&(q_\lambda q_\mu -q^2 g_{\lambda\mu})-\xi q_\lambda q_\mu
+\Pi_{\lambda\mu}(q).\label{Dbaremoinsun}
\end{eqnarray}
The vacuum polarization
tensor ${\Pi}_{\lambda\mu}(q)$ is transverse \cite{Itzykson}.
If we multiply (\ref{Dbaremoinsun}) by (\ref{DT}),
we obtain
\begin{eqnarray}
g_\lambda^{\,\,\nu} - \frac{q_\lambda q^\nu}{q^2} &=&
\big((q_\lambda q_\mu -q^2 g_{\lambda\mu})+{\Pi}_{\lambda\mu}(q)\big)
{D}^{T\mu\nu}(q).\label{bareproj}
\end{eqnarray}
The left-hand side of Eq.(\ref{bareproj}) is the projector
onto the transverse tensors.

We show in section \ref{phipsi} that
\begin{eqnarray}
\Pi^{\lambda\mu}(q)=i e_0^2
  \int \frac{\dd^4p}{(2\pi)^4}
  \mathrm{tr}\big[\gamma^\lambda
  \frac{\delta S(p)}{e_0 \delta A^0_\mu(-q)}\big]. \label{Piphi1}
\end{eqnarray}

\subsection{Renormalized propagators}
To obtain the Schwinger equations for the renormalized propagators,
the best is to start from the renormalized Lagrangian, and
to follow the steps given by Bogoliubov and Shirkov
\cite{Bogoliubov}, Itzykson and Zuber \cite{Itzykson}
or Rochev \cite{Rochev}. However, to give an idea of the
result, we introduce some of Dyson's recipes.

The longitudinal part of the photon Green function is 
not modified by renormalization \cite{Itzykson}, 
and the renormalized photon propagator can be decomposed as
\begin{eqnarray}
{\bar D}_{\lambda\mu}(q)&=&{\bar D}^T_{\lambda\mu}(q)-
\frac{1}{\xi} \frac{q_\lambda q_\mu}{(q^2+i\epsilon)^2},\label{DbarT}
\end{eqnarray}
where ${\bar D}^T_{\lambda\mu}(q)$ is transverse.

Then, we introduce Dyson's relation between renormalized and bare
Green functions (Ref.\cite{Itzykson} p.414):
\begin{eqnarray}
{\bar S}(q)Z_2 &=& S(q),\label{DysonS}\\
Z_3{\bar D}^T_{\mu\nu}(q) &=& D^T_{\mu\nu}(q),\label{DysonDT}\\
Z_3 e_0^2 &=& e^2,\label{Ward}\\
m_0 &=& m-\delta m,\label{Dysonm}
\end{eqnarray}
where $Z_2$ and $Z_3$ are (infinite) scalars independent of $q$,
and $e$ is the renormalized charge.
Equation (\ref{Ward}) was conjectured by
Dyson \cite{Dyson} and proved by Ward (\cite{Itzykson}, p.413).
Finally, the external field $A^0_\lambda$ is renormalized as 
$A_\lambda$, so that 
\begin{eqnarray}
e_0A^0_\lambda &=& eA_\lambda.\label{extfield}
\end{eqnarray}

To introduce the mass renormalization, we must
start from the differential form of the Schwinger
equation for the bare electron propagator, where
we reintroduce the external field, for later convenience,
\begin{eqnarray}
\big[i\gamma\cdot\partial-m_0-e_0\gamma\cdot A^0(x)] S(x,y;A^0)
 &=&\delta(x-y)+
\nonumber\\&& \hspace*{-45mm}
 ie_0^2\int d^4z\,\gamma^\mu
  D_{\mu\rho}(x,z;A^0)\frac{\delta S(x,y;A^0)}{\delta
e_0A^0_{\rho}(z)}. \label{IZ2}
\end{eqnarray}

Dyson showed that relations (\ref{DysonS}-\ref{extfield}),
are valid in the presence of an external field.
We use them in Eq.(\ref{IZ2}) to obtain
\begin{eqnarray}
\big[i\gamma\cdot\partial-m] \bar S(x,y;A)Z_2
 &=&\delta(x-y)-\delta m S(x,y;A)Z_2
\nonumber\\&& \hspace*{-35mm}
+ie^2\int d^4z\,\gamma^\mu
  \bar D_{\mu\rho}(x,z;A)\frac{\delta S(x,y;A)}{\delta
eA_{\rho}(z)}Z_2. \label{RIZ2}
\end{eqnarray}
In Eq.(\ref{RIZ2}), we have changed the gauge parameter
$\xi_0$ of $D_{\mu\rho}(x,z;A^0)$ into $\xi=Z_3\xi_0$
(see Ref.\cite{Itzykson} p. 414).

If we multiply Eq.(\ref{RIZ2}) by 
\begin{eqnarray}
S^0(z,y;A)=[i\gamma\cdot\partial-m-e\gamma\cdot A]^{-1}
\label{S0A}
\end{eqnarray}
and integrate over $x$ we obtain
the integral Schwinger equation for the renormalized 
electron propagator,
\begin{eqnarray}
\bar S(x,y;A) Z_2 &=& S^0(x,y;A)
\nonumber\\&& \hspace*{-21mm}
+ie^2\int d^4z d^4z' S^0(x,z;A)
  \gamma^\lambda \bar D_{\lambda\lambda'}(z,z';A)
  \frac{\delta \bar S(z,y;A)}{e\delta A_{\lambda'}(z')}Z_2
\nonumber\\&& \hspace*{-10mm}
-\delta m \int d^4z S^0(x,z;A) \bar S(z,y;A) Z_2.
\label{eqbasexy}
\end{eqnarray}

In Eq.(\ref{eqbasexy}), we put $A=0$ and we Fourier
transform to find
\begin{eqnarray}
{\bar S}(q)Z_2 &=& S^0(q)-
  \delta m S^0(q){\bar S}(q) Z_2
 \nonumber\\&&\hspace*{-8mm}+
 ie^2 S^0(q)
  \int\frac{\dd^4p}{(2\pi)^4}
  \gamma^\lambda {\bar D}_{\lambda\lambda'}(p)
  \frac{\delta {\bar S}(q-p)}{e\delta A_{\lambda'}(p)}Z_2.
\label{eqbaseelectron}
\end{eqnarray}

This equations was given by Bogoliubov and Shirkov
\cite{Bogoliubov}, as well as by Itzykson and Zuber
(\cite{Itzykson}, p.481), except for the mass
counterterm $\delta m$ which was apparently overlooked
by these authors. A complete derivation can be
found in \cite{Rochev} (notice that his
$\delta m$ is our $Z_2 \delta m$). 

To obtain a convenient Schwinger equation for
the renormalized photon propagator, we 
must introduce the 
renormalized vacuum polarization
$\bar \Pi_{\lambda\mu}(q)$, defined by
\begin{eqnarray}
{[{\bar D}^{-1}]}_{\lambda\mu}(q) & = & 
(q_\lambda q_\mu -q^2 g_{\lambda\mu})-\xi q_\lambda q_\mu
+{\bar\Pi}_{\lambda\mu}(q).\label{Dmoinsun}
\end{eqnarray}

It may be useful to compare these definitions to those of Itzykson
and Zuber \cite{Itzykson}: $\bar{D}_{\mu\nu}=-i \bar{G}_{\mu\nu}$,
$\bar{\Pi}_{\mu\nu}=-i \bar\omega_{\mu\nu}$.

If we multiply (\ref{Dmoinsun}) by (\ref{DbarT}),
we obtain
\begin{eqnarray}
g_\lambda^{\,\,\nu} - \frac{q_\lambda q^\nu}{q^2} &=&
\big((q_\lambda q_\mu -q^2 g_{\lambda\mu})+{\bar\Pi}_{\lambda\mu}(q)\big)
{\bar D}^{T\mu\nu}(q).\label{renproj}
\end{eqnarray}

If we compare Eqs.(\ref{bareproj}) and (\ref{renproj}),
and use (\ref{DysonDT}), we find
\begin{eqnarray}
q_\lambda q_\mu -q^2 g_{\lambda\mu}+{\bar\Pi}_{\lambda\mu}(q) &=&
Z_3(q_\lambda q_\mu -q^2 g_{\lambda\mu} \nonumber\\&& +\Pi_{\lambda\mu}(q)).
\label{PiZtrois}
\end{eqnarray}

Therefore, using Eqs. (\ref{DysonS}) and (\ref{extfield})
\begin{eqnarray*}
\Pi^{\lambda\mu}(q)=i e_0^2 Z_2
 \int \frac{\dd^4p}{(2\pi)^4}
  \mathrm{tr}\big[\gamma^\lambda
  \frac{\delta {\bar S}(p)}{e \delta A_\mu(-q)}\big].
\end{eqnarray*}
Introducing this equation into (\ref{PiZtrois}),
and using Eq.(\ref{Ward}) we obtain

\begin{eqnarray}
{\bar\Pi}_{\lambda\mu}(q) &= &(Z_3-1)(q_\lambda q_\mu -q^2 g_{\lambda\mu})
\nonumber\\
&&+ Z_2 ie^2 \int \frac{\dd^4p}{(2\pi)^4}
  \mathrm{tr}\big[\gamma_\lambda
  \frac{\delta {\bar S}(p)}{e \delta A^\mu(-q)}\big].
\label{eqbasephoton}
\end{eqnarray}

Equations (\ref{eqbaseelectron}) and
(\ref{eqbasephoton}) will be the bases of
recursive expressions for the renormalized
electron and photon propagators.

\section{Tree expansion of propagators}
For the convenience of the reader, and because the
notation of Ref. \cite{BrouderEPJC2} as been modified
\footnote{There are no longer
trees with black or white roots. The color of
the root is now indicated by the function $\varphi$
itself. $\varphi(t;q)$ corresponds to a tree
with a black root, $\varphi_{\mu\nu}(t;q)$
to a tree with a white root.
This notation is more elegant than the
one used in \cite{BrouderEPJC2}.},
we recall the description of photon and electron propagators
in terms of planar binary trees.

\subsection{Trees and propagators}
The main trick of Ref.\cite{BrouderEPJC2} was to write 
each propagator as a sum indexed by planar binary trees,
to be defined in the next section.

The bare electron Green function in Fourier space,
$S(q)$ is written as
a sum over planar binary trees $t$
\begin{eqnarray}
S(q)&=& \sum_{t} e_0^{2|t|} \varphi^0 (t;q). \label{bareS}
\end{eqnarray}
Here $e_0$ is the bare electron charge (i.e. the electron
charge before renormalization). The fact that
the expansion is over $e_0^2$ (and not $e_0$)
was justified in Ref.\cite{BrouderEPJC2}.
Similarly, the renormalized electron Green function
is expanded over planar binary trees
\begin{eqnarray}
\bar S(q)&=& \sum_{t} e^{2|t|} \bar\varphi^0 (t;q). \label{renS}
\end{eqnarray}
In Eq.(\ref{renS}) $e$ is the renormalized (finite)
electron charge.

The bare and renormalized photon Green functions
are written as
\begin{eqnarray}
D_{\mu\nu}(q)&=& \sum_{t} e_0^{2|t|} \varphi_{\mu\nu}^0 (t;q),\nonumber\\
\bar D_{\mu\nu}(q)&=& \sum_{t} e^{2|t|} \bar\varphi_{\mu\nu}^0 (t;q).
\label{renD}
\end{eqnarray}

For the renormalization of the photon Green function and
the vacuum polarization, it will be necessary to distinguish
the photon Green function $D_{\mu\nu}(q)$ and the 
transverse photon Green function $D^T_{\mu\nu}(q)$. Since all terms
$\varphi_{\mu\nu}^0 (t;q)$ and
$\bar\varphi_{\mu\nu}^0 (t;q)$ are transverse for $t\not=\|$,
the transverse renormalized propagator is
\begin{eqnarray*}
\bar D^T_{\mu\nu}(q)&=&\varphi^T_{\mu\nu}(\|;q)
+\sum_{|t|>0} e^{2|t|} \bar\varphi_{\mu\nu}^0 (t;q),
\end{eqnarray*}
where $ \varphi^T_{\mu\nu}(\|;q)=D^{0T}_{\mu\nu}(q)$.

The bare and renormalized vacuum polarization are
expanded similarly:
\begin{eqnarray}
\Pi_{\lambda\mu}(q)
& = & \sum_{|t|>0} e_0^{2|t|} \varpsi^0_{\lambda\mu}(t;q),
\label{defPi}\\
\bar\Pi_{\lambda\mu}(q)
& = & \sum_{|t|>0} e^{2|t|} \bar\varpsi^0_{\lambda\mu}(t;q).
\label{defPibar}
\end{eqnarray}
For later convenience, we finally define
\begin{eqnarray*}
\varpsi^0_{\lambda\mu}(\|;q) = q_{\lambda} q_\mu -q^2
g_{\lambda\mu},
\end{eqnarray*}
so that
\begin{eqnarray}
{[{D^0}^{-1}]}_{\lambda\mu}(q) & = & (q_\lambda q_\mu -q^2 g_{\lambda\mu})
-\xi q_\lambda q_\mu\\
& = & \varpsi^0_{\lambda\mu}(\|;q) -\xi q_\lambda q_\mu.
\label{defDmun}
\end{eqnarray}

A few identities will be useful in the sequel
\begin{eqnarray*}
D^0_{\lambda\mu}(q) (q^{\mu} q_\nu -q^2 g^\mu_{\,\,\nu})&=&
-q^2 D^{0T}_{\lambda\nu}(q),\\
\varpsi_{\lambda\lambda'}^0(\|;q)
D^{0\lambda'\mu'}(q)\varpsi_{\mu'\mu}^0(\|;q)&=&
\varpsi_{\lambda\mu}^0(\|;q),\\
D^{0}_{\lambda\lambda'}(q)\varpsi^{0\lambda'\mu'}(\|;q)
D^{0}_{\mu'\mu}(q)&=& D^{0T}_{\lambda\mu}(q).
\end{eqnarray*}
Notice that $-q^2 D^{0T}_{\lambda\nu}(q)$ is the
projector onto the transverse tensors.

The tree representation of the photon propagator
enjoys the following property \cite{BrouderEPJC2}
\begin{eqnarray*}
\varphi^0_{\mu\nu}(t_l\vee t_r;q)&=&
\varphi^0_{\mu\lambda}(\|\vee t_r;q)
[(D^0)^{-1}]^{\lambda\lambda'}(q)\,
\varphi^{0}_{\lambda'\nu}(t_l;q).
\end{eqnarray*}

This equality is non trivial only if $t_l\not= \|$. But
then $\varphi^0_{\mu\lambda}(\|\vee t_r;q)$ and
$\varphi^{0}_{\lambda'\nu}(t_l;q)$ are transverse 
\cite{BrouderEPJC2}, this cancels the term proportional
to $\xi$ and we can write
\begin{eqnarray}
\varphi^0_{\mu\nu}(t_l\vee t_r;q)&=&
\varphi^0_{\mu\lambda}(\|\vee t_r;q)
\varpsi^{0\lambda\lambda'}(\|;q)
\varphi^{0}_{\lambda'\nu}(t_l;q). \label{phiphi2}
\end{eqnarray}
The fact that $\varphi^0_{\mu\nu}(t_l\vee t_r;q)$
is a product for $t_l\not= \|$ can be checked
in appendix 3.

As shown in section \ref{phipsi}, 
$\varpsi^0_{\mu\nu}(t_l\vee t_r)=0$ if $t_l\not= \|$ and 
\begin{eqnarray}
\varpsi^0_{\mu\nu}(\|\vee t_r;q)&=&
-\varpsi^0_{\mu\lambda}(\|;q)
\varphi^{0\lambda\lambda'}(\|\vee t_r;q)
\varpsi^0_{\lambda'\nu}(\|;q). \label{psiphipsi}
\end{eqnarray}

\subsection{Trees and renormalization constants}

According to Dyson's multiplicative renormalization
of QED, we define three renormalization constants
$Z_2$, $Z_3$ and $\delta m$. We expand these constants
over planar binary trees
\begin{eqnarray}
Z_2 &=& \sum_t e^{2|t|}\zeta_2(t),
  \quad\mathrm{with}\quad\zeta_2(\|)=1, \label{treeZ2}\\
Z_3 &=& \sum_t e^{2|t|}\zeta_3(t),
  \quad\mathrm{with}\quad\zeta_3(\|)=1, \label{treeZ3}\\
\delta m &=& \sum_t e^{2|t|}\zeta_m(t),
  \quad\mathrm{with}\quad\zeta_m(\|)=0.\label{treeZm}
\end{eqnarray}

These expansions will be the basis of the tree by tree
renormalization of QED.

\subsection{Planar binary trees}
A planar binary tree is a tree with a designated
vertex called the root. To follow the notation of
Loday and Ronco \cite{LodayRonco}, we write
the root vertex as $\|$. The other vertices are
not explicitly drawn, but they are at the ends
of each edge, which are $\\$ or $/$.
The trees are binary because each
vertex has either zero or two children. They
are planar because $\deuxun$ is different
from $\deuxdeux$. 
The planar binary trees have an odd number of vertices
and for each tree $t$ we define $|t|$ as the integer such
that $t$ has $2|t|+1$ vertices. In other words, $|t|$ is
the number of internal vertices.
We call $Y_n$ the set of planar binary trees $t$ with 
such that $|t|=n$.

The planar binary trees with up to 7 vertices are
\begin{eqnarray*}
Y_0 &=& \{\|\},\\
Y_1 &=& \{\Y\},\\
Y_2 &=& \{\deuxun,\,\deuxdeux\},\\
Y_3 &=& \{\troisun,\,\troisdeux,\,\troistrois,
  \,\troisquatre,\,\troiscinq\}.
\end{eqnarray*}

We denote $Y$ the set of all planar binary trees
\begin{eqnarray*}
Y=\overset{\infty}{\underset{n=0}{\bigcup}} Y_n.
\end{eqnarray*}

Finally we consider the operation of {\em grafting\/} two trees, 
$\vee : Y_p \times Y_q \longrightarrow Y_{p+q+1}$, 
by which the roots of two trees $t_1$ and $t_2$ are joined into a new 
vertex that becomes the root of the tree
$t = t_1 \vee t_2$, cf \cite{LodayRonco}. For instance
\begin{eqnarray}
\Y\vee \Y &=& \troistrois.
\end{eqnarray} 
It is clear that any tree $t$, 
except the $0$-tree $\|$, is the grafting of two uniquely determined trees
$t_l$ and $t_r$ with orders $|t_l|,|t_r| \leq |t|-1$. 

\subsection{Recursive equations for bare propagators 
\label{recursection}}
In Ref.\cite{BrouderEPJC2}, we have obtained recursive relations
for $\varphi(t)$ and $\varphi_{\lambda\mu}(t)$.

For the electron propagator,
$\varphi(t)$ satisfies the recursive relation
\begin{eqnarray}
\varphi^n(t;q;\{\lambda,p\}_{1,n})&=&
 S^0(q)\gamma^{\lambda_1}\nonumber\\&&\hspace*{-23mm}
 \times\varphi^{n-1}(t;q+p_1;\{\lambda,p\}_{2,n})
  \nonumber\\&&\hspace*{-23mm}
 +i\sum_{k=0}^{n}\int \frac{\dd^4 p}{(2\pi)^4} S^0(q)\gamma^\lambda
 \varphi^k_{\lambda\lambda'} (t_l;p;\{\lambda,p\}_{1,k})
  \nonumber\\&&\hspace*{-23mm}
 \times\varphi^{n-k+1}_\Sigma(t_r;q-p;\lambda',p+P_k,
  \{\lambda,p\}_{k+1,n}),\label{phinoirk}
\end{eqnarray}
where we have noted $P_k=p_1+\cdots+p_k$, ($P_0=0$) and
$\{\lambda,p\}_{1,n}={\lambda_1,p_1,\dots,\lambda_n,p_n}$.
The initial data are
\begin{eqnarray}
\varphi^0(\|;q) &=& S^0(q),\nonumber\\
\varphi^1(\|;q;\lambda_1,p_1) &=&
S^0(q)\gamma^{\lambda_1}S^0(q+p_1),\nonumber\\
\varphi^n(\|;q;\{\lambda,p\}_{1,n}) &=&
S^0(q)\gamma^{\lambda_1}
  S^0(q+p_1)\gamma^{\lambda_2}\cdots\gamma^{\lambda_n}\nonumber\\&&
  \hspace*{5mm}
  S^0(q+p_1+\cdots+p_n). \label{initcond}
\end{eqnarray}

The symbol  $\varphi^{n+1}_\Sigma(t;q;z,\{z\}_{1,n})$
is defined as the sum of $n$ terms,
where the first variable $z=(\lambda,p)$ is exchanged in turn
with all the variables $z_i=(\lambda_i,p_i)$.
\begin{eqnarray*}
\varphi^{n+1}_\Sigma(t;q;z,\{z\}_{1,n}) &=&
\varphi^{n+1}(t;q;z,\{z\}_{1,n})  \\&&\hspace*{-25mm}+
\varphi^{n+1}(z_1,z,\{z\}_{2,n})+\cdots+
\varphi^{n+1}(\{z\}_{1,n-1},z,z_n) \\&&+\varphi^{n+1}(\{z\}_{1,n},z).
\end{eqnarray*}

For the photon propagator,
$\varphi_{\mu\nu}(t)$ satisfies the recursive relation
\begin{eqnarray}
\varphi^n_{\mu\nu}(t;q;\{\lambda, p\}_{1,n})&=&
 -i\sum_{k=0}^{n}\int \frac{\dd^4p}{(2\pi)^4} D^0_{\mu\lambda}(q)
  \nonumber\\&&\hspace*{-29mm}
  \times\mathrm{tr}\big[\gamma^\lambda
  \varphi^{k+1}_\Sigma(t_r;p;\lambda',-q-P_{k},
   \{\lambda, p\}_{1,k})
  \big]
  \nonumber\\&&\hspace*{-29mm}
 \times\varphi^{n-k}_{\lambda'\nu}(t_l;q+P_{k};
  \{\lambda, p\}_{k+1,n}),
  \label{phiblanck}
\end{eqnarray}
with the initial data
\begin{eqnarray*}
\varphi^0_{\mu\nu}(\|;q) &=& D^0_{\mu\nu}(q),\\
\varphi^n_{\mu\nu}(\|;q;\{\lambda, p\}_{1,n}) &=& 
  0\quad\mathrm{for}\quad n\ge 1.
\end{eqnarray*}

In this paper, we use a non symmetrized 
$\varphi^n(\|)$. However, for the validity of 
Ward-Takahashi identities, it would be necessary
to handle symmetrized expressions
\begin{eqnarray*}
\frac{1}{n!}\sum_{\sigma\in {\cal S}_n}
\varphi^n(\|;q;\lambda_{\sigma(1)},p_{\sigma(1)},\dots,
\lambda_{\sigma(n)},p_{\sigma(n)}).
\end{eqnarray*}
Such expressions seem complicated. It is possible
that recent string-like methods help handling them
\cite{Lam,Schmidt}.

\subsection{The pruning operator}

In this section, we introduce the pruning operator $P$
which will prove very useful to obtain a recursive
expression for renormalized propagators. If $t$
is a tree, $P(t)$ is a sum of $n(t)$ terms of the form
$u_j \otimes v_j$, where $u_j$ and $v_j$ are planar
binary trees. More mathematically
\begin{eqnarray}
P(t)=\sum_{j=1}^{n(t)} u_j \otimes v_j. \label{Pdet}
\end{eqnarray}
Before we fully define $P(t)$, we want to show why
it is useful. If, for each tree $t$, $\varphi(t)$
and $\psi(t)$ are $4\times 4$ complex matrices, we
call the convolution of $\varphi$ and $\psi$ the quantity
\begin{eqnarray*}
(\varphi\star\varpsi)(t)&=&\sum_{i=1}^{n(t)} \varphi(u_i)\varpsi(v_i). 
\label{convolution}
\end{eqnarray*}
The main property of this convolution was established
in Ref.\cite{BrouderEPJC2}. If
\begin{eqnarray*}
X(\lambda) &=& \sum_{t} \lambda^{|t|} x(t)\quad\mathrm{and}\quad
Y(\lambda) = \sum_{t} \lambda^{|t|} y(t),
\end{eqnarray*}
with $x(\|)=y(\|)=0$, then
\begin{eqnarray*}
X(\lambda)Y(\lambda) &=& \sum_{t} \lambda^{|t|} (x\star y)(t).
\end{eqnarray*}
In other words, the pruning operator and the convolution
enable us to multiply series indexed by planar binary trees.

This nice property justifies the trouble of introducing
$P(t)$. First, $n(t)$, the number of terms in Eq.(\ref{Pdet}),
is defined by $n(\|)=0$ and
\begin{eqnarray}
n(t) &= & 0\quad\mathrm{if}\quad t=t_l\vee\|, \nonumber\\
n(t) &= & n(t_r)+1
 \quad\mathrm{if}\quad t=t_l\vee t_r,\quad t_r\not=\|.
\end{eqnarray}
Finally, $P(t)$ is determined recursively by
$P(\|)=0$ and
\begin{eqnarray}
P(t) &= & 0\quad\mathrm{if}\quad t=t_l\vee\|, \nonumber\\
P(t) &= & (t_l\vee \|)\otimes t_r+
  \sum_{j=1}^{n(t_r)} (t_l\vee u_j)\otimes
v_j\nonumber\\&&\hspace*{8mm}
 \quad\mathrm{if}\quad t=t_l\vee t_r,\quad t_r\not=\|.
 \label{defP}
\end{eqnarray}
The trees $u_j$ and $v_j$ in Eq.(\ref{defP}) are generated by
Eq.(\ref{Pdet}) for $t=t_r$.
For instance
\begin{eqnarray*}
P(\|) = P(\Y) = &P(\deuxun) &= P(\troisun) = P(\troisdeux) = 0, \\
P(\deuxdeux) = \Y \otimes \Y, &&
P(\troistrois)= \deuxun \otimes \Y, \\
P(\troisquatre)= \Y \otimes \deuxun, &&
P(\troiscinq)= \Y \otimes \deuxdeux + \deuxdeux \otimes \Y .
\end{eqnarray*}
We show in the appendix that the pruning operator is
coassociative, that is
\begin{eqnarray}
\label{uffa}
(P \otimes id) \otimes P = (id \otimes P) \otimes P. 
\end{eqnarray}
Therefore the convolution is associative.

We consider on trees the structure of an associative algebra
$T(Y)$ given by the (non commutative) tensor product,
$T(Y) = Y \oplus Y^{\otimes 2} \oplus Y^{\otimes 3} \oplus \ldots$,
and we set the root $\|$ as the unit:
$\| \otimes t = t \otimes \| = t$.
Then we extend $P$ to $T(Y)$ as a multiplicative map,
but $P$ does not preserve the unit, since
$P(\|)$ is not equal to $\| \otimes \|$. 
We can define a coproduct
\begin{eqnarray*}
\DeltaP t &=& \|\otimes t + P(t) + t \otimes \|,\\
\DeltaP \| &=& \|\otimes \|.
\end{eqnarray*}

This $\DeltaP$ is the coproduct of a Hopf algebra
over planar binary trees.
Its antipode is given by the recursive formula 
\begin{eqnarray}
\Sstar(t) &=& -t - (\Id\star \Sstar)(t) =
-t - (\Sstar \star \Id)(t), \label{defantipode}
\end{eqnarray}
for $t\not= \|$, and $\Sstar(\|)=\|$.

To define the convolution of $x(t)$ and $y(t)$,
we needed the condition 
$x(\|)=y(\|)=0$. When this condition
is not satisfied, we have two solutions.
The first solution is to isolate the root, so that
\begin{eqnarray*}
X(\lambda)Y(\lambda) &=& 
x(\|)y(\|)+(X(\lambda)-x(\|))y(\|)\\&&\hspace*{-10mm}
+x(\|)(Y(\lambda)-y(\|)
+(X(\lambda)-x(\|))(Y(\lambda)-y(\|))\\
&=& x(\|)y(\|)+ x(\|)\sum_{|t|>0} \lambda^{|t|} y(t)\\&&
   +\sum_{|t|>0} \lambda^{|t|} x(t) y(\|) +
\sum_{|t|>0} \lambda^{|t|} (x\star y)(t).
\end{eqnarray*}

The second solution is to use the coproduct $\DeltaP$. 
Thus, we define the convolution $\nablaP$ by:
\begin{eqnarray*}
(x\nablaP y)(t)=\sum_j x(u_j)y(v_j),\,\,\mathrm{where}\,\,
\Delta^P(t)=\sum_j u_j\otimes v_j.
\end{eqnarray*}

With this alternative convolution, the equality
\begin{eqnarray}
X(\lambda)Y(\lambda) &=& 
   \sum_{t} \lambda^{|t|} (x\nablaP y)(t)
\label{mulnablaP}
\end{eqnarray}
is satisfied even if $x$ or $y$ is not zero on the root.

In our final formulas, we prefer to use the convolution
$\star$ because its ensures the recursivity of the 
expressions (the trees in $(x\star y)(t)$ are strictly
smaller than $t$).

A last point of notation. If $\varphi$ and $\varpsi$
depend on other arguments, we leave them inside
$\varphi$ and $\varpsi$. For example
\begin{eqnarray*}
(\varphi(q)\star\varpsi(q))(t)=\sum_{j=1}^{n(t)} \varphi(u_j;q)
  \varpsi(v_j;q).
\end{eqnarray*}

\subsection{The self-energy}

As a first application of the convolution defined in the
previous section, we introduce the tree-expansion for the
electron self-energy.

The bare electron self-energy $\Sigma(q)$ is defined by:
\begin{eqnarray*}
S^{-1}(q)&=& \gamma\cdot q -m -\Sigma(q)
= -\sum_{t} e_0^{2|t|} \varpsi^0(t;q),
\end{eqnarray*}
where
\begin{eqnarray*}
\varpsi^0(\|;q)&=& -(\gamma\cdot q -m)\quad\mathrm{and}\quad
\Sigma(q)= \sum_{|t|>0} e_0^{2|t|} \varpsi^0(t;q),
\end{eqnarray*}
so that $\varpsi^0(\|;q)\varphi^0(\|;q)=-1$.
The pruning operator is used to define the expansion
of the bare self-energy over trees in terms of the
expansion of the bare electron Green function over trees:
\begin{eqnarray}
\varpsi^0(t)&=& \varpsi^0(\|)\varphi^0(t)\varpsi^0(\|)
+\varpsi^0(\|)(\varphi^0\star\varpsi^0)(t), \label{inversepsi}\\
\varphi^0(t)&=& \varphi^0(\|)\varpsi^0(t)\varphi^0(\|)
+(\varphi^0\star\varpsi^0)(t)\varphi^0(\|) \label{inversephi}.
\end{eqnarray}
In terms of the antipode, (\ref{inversepsi}) can be rewritten
\begin{eqnarray}
\varpsi^0(t)=-\varpsi^0(\|;q)(\varphi^0(q)\circ \Sstar)(t)
\varpsi^0(\|;q).\label{antipobare}
\end{eqnarray}
Similarly, for the renormalized self-energy, we have
\begin{eqnarray}
\bar\varpsi^0(t)=-\varpsi^0(\|;q)(\bar\varphi^0(q)\circ \Sstar)(t)
\varpsi^0(\|;q).\label{antiporen}
\end{eqnarray}
We must give some detail concerning the meaning of expressions
like $(\varphi^0(q)\circ \Sstar)(t)$. Because of its definition
(\ref{defantipode}), the antipode $\Sstar$ acting on $t$
generates a sum of products of trees.
The action of
$\varphi^0(q)$ on this sum is prolonged
from its action on $Y$ to an algebra homomorphism
over $T(Y)$. In other words
\begin{eqnarray*}
\varphi^0(q)(t_1+t_2) & = & \varphi^0(t_1;q)+\varphi^0(t_2;q),\\
\varphi^0(q)(\lambda t) & = & \lambda \varphi^0(t;q).
\end{eqnarray*}
For a product of trees, we do not want to simply
multiply two Feynman diagrams for the electron propagator,
we must cancel one of the free propators between them.
Thus, the product is 
\begin{eqnarray}
\varphi^0(q)(t_1 t_2) & = & -\varphi^0(t_1;q)\varpsi^0(\|;q)
\varphi^0(t_2;q). \label{produitn}
\end{eqnarray}
This operation becomes clear if one tries it on some examples
given in appendix 3. Since the (matrix) product on the right-hand side
of Eq.(\ref{produitn}) is not commutative, the algebra
product on trees is not commutative either.

In the presence of an external field $A$, the definition
(\ref{S0A}) of $S^0(z,y;A)$ gives, after inversion and
Fourier transform, 
\begin{eqnarray}
\varpsi^0(\|;q;A) &=& -(\gamma\cdot q -m-e\gamma\cdot A(q) )
\label{varpsi0}
\end{eqnarray}
Thus we obtain, at $A^0=0$
\begin{eqnarray*}
\varpsi^0(\|;q)&=& -(\gamma^\alpha q_\alpha -m),\\
\varpsi^1(\|;q;\lambda,p)&=& \gamma^\lambda,\\
\varpsi^n(\|;q;\{\lambda,p\}_{1,n}) &=& 0\quad\mathrm{for}
\quad n>1.
\end{eqnarray*}

The components of $\varpsi^0(t;q)$ for the other trees
$t$ are obtained by using the chain rule for the
functional derivative of (\ref{inversepsi}) with
respect to $e_0 A^0_{\lambda_i}(p_i)$, taken at $A^0=0$.
For $n=1$, this gives the same result as in sect. 6.4 of
\cite{BrouderEPJC2}.

Finally it is shown in section \ref{deuxg}
that the bare self-energy can
be calculated from the recursive equation
\begin{eqnarray}
\varpsi^0(t;q) &=&
 i \int \frac{\dd^4p}{(2\pi)^4}\gamma^{\lambda}
  \varphi^0_{\lambda\lambda'}(t_l;p)
  g(t_r;q-p;\lambda',p), \label{intpsizero}
\end{eqnarray}
where
\begin{eqnarray*}
g(t_r;q-p;\lambda',p) &=& -(\varphi^1(q-p;\lambda',p)\nablaP\varpsi^0(q))(t_r)\\
&=&
(\varphi^0(q)\nablaP\varpsi^1(q-p;\lambda',p))(t_r).
\end{eqnarray*}

\subsection{The higher components $\varphi^n(t)$ \label{compon}}

For a complete recursive solution of the renormalized
propagators, we must define the higher components
$\bar\varphi^n(t)$ and $\bar\varphi^n_{\mu\nu}(t)$.
As for the bare propagators (see Ref. \cite{BrouderEPJC2}), they
are defined as the functional derivative with respect to an external
electromagnetic field. 

Let us be more accurate concerning this
external field. As noticed by Bogoliubov and Shirkov, 
Eq.(\ref{eqbasexy}) is not the Schwinger equation for QED with
an external electromagnetic field, since the latter involves
tadpole diagrams which are absent from Eq.(\ref{eqbasexy}).
However, Eq.(\ref{eqbasexy}) is the Schwinger equation for
a renormalizable theory (i.e. QED without tadpoles), and 
Dyson's relations (\ref{DysonS}-\ref{extfield}) still hold.

In the real space, the bare and renormalized electron Green functions
are expanded as
\begin{eqnarray*}
S(x,y;A) &=& \sum_t e_0^{2|t|} \varphi^0(t;x,y;A),\\
\bar{S}(x,y;A) &=& \sum_t e^{2|t|} \bar\varphi^0(t;x,y;A).
\end{eqnarray*}
In these expressions, we do not distinguish between $A$ and $A^0$
because $A^0$ comes always multiplied by $e_0$ and
$e_0 A^0=eA$. On the root, we have 
$\varphi^0(\|;x,y;A)=S^0(x,y;A)$. 

The higher components of $\bar\varphi^0(t;x,y;A)$ must satisfy
\begin{eqnarray*}
\frac{\delta}{e\delta A_\lambda(z)} \varphi^n(t;x,y;\{\lambda,z\}_{1,n};A)
&=& \\&&\hspace*{-20mm}
\varphi^{n+1}_\Sigma(t;x,y;\lambda,z,\{\lambda,z\}_{1,n};A).
\end{eqnarray*}
where the notation $\varphi^{n+1}_\Sigma$ and $\{\lambda,z\}_{1,n}$ 
is defined in section \ref{recursection}.

Since our purpose is QED without external field, $A$ is just
used to take functional derivatives, and the higher components
we actually need are
\begin{eqnarray*}
\varphi^n(t;x,y;\{\lambda,z\}_{1,n})=\varphi^n(t;x,y;\{\lambda,z\}_{1,n};A)
\end{eqnarray*}
for $A=0$.

At $A=0$, the theory becomes translational invariant, and
a Fourier transform gives us
\begin{eqnarray*}
\frac{\delta}{e\delta A_\lambda(p)} \varphi^n(t;q;\{\lambda,p\}_{1,n})
&=& \varphi^{n+1}_\Sigma(t;q;\lambda,p,\{\lambda,p\}_{1,n}).
\end{eqnarray*}

In the recursive equations for $\varphi(t)$ we meet products
of propagators such as $\varphi^0(t_1;q)\varphi^0(t_2;q)$.
In the real space, this gives a convolution of 
$\varphi^0(t_1;x,y)$ and $\varphi^0(t_2;x,y)$. We take the
functional derivative of this convolution with respect to
$A(z)$, we Fourier transform the result and we obtain, at $A=0$
\begin{eqnarray*}
\frac{\delta}{e\delta A_\lambda(p)} \big(\varphi^0(t_1;q)
\varphi^0(t_2;q) \big) &=& \varphi^0(t_1;q)\varphi^1(t_2;q;\lambda,p)
\\&&\hspace*{-5mm}
+\varphi^1(t_1;q;\lambda,p)\varphi^0(t_2;q+p).
\end{eqnarray*}
This expression satisfies energy-momentum conservation.

To take the functional derivatives of the recursive
equations for renormalized quantities, we need 
the independence of the renormalization constants with
respect to the external field. 
There are various ways to prove this. For instance,
the differential form of the Ward identity (Eq.(21)
in Ref.\cite{BrouderEPJC2} is
\begin{eqnarray*}
\frac{\partial\varphi^n(t;q;\{\lambda,p\}_{1,n})}{\partial q^\mu}=
-\varphi^{n+1}_\Sigma(t;q;\mu,0,\{\lambda,p\}_{1,n}).
\end{eqnarray*}
The Ward identity is also valid for the renormalized electron
propagator \cite{Jauch}, so that
\begin{eqnarray*}
\frac{\partial\bar\varphi^n(t;q;\{\lambda,p\}_{1,n})}{\partial q^\mu}=
-\bar\varphi^{n+1}_\Sigma(t;q;\mu,0,\{\lambda,p\}_{1,n}).
\end{eqnarray*}
From the definitions (\ref{antipobare}) and (\ref{antiporen})
of the bare and renormalized self-energies we obtain
\begin{eqnarray*}
\frac{\partial\varpsi^n(t;q;\{\lambda,p\}_{1,n})}{\partial q^\mu} &=&
-\varpsi^{n+1}_\Sigma(t;q;\mu,0,\{\lambda,p\}_{1,n}),\\
\frac{\partial\bar\varpsi^n(t;q;\{\lambda,p\}_{1,n})}{\partial q^\mu} &=&
-\bar\varpsi^{n+1}_\Sigma(t;q;\mu,0,\{\lambda,p\}_{1,n}).
\end{eqnarray*}

Now we start from the relation between the renormalized and
bare self-energies, for instance
\begin{eqnarray}
\bar\varpsi^0(\Y;q) &=& \varpsi^0(\Y;q) - \zeta_2(\Y) (\gamma\cdot q -m)
  -\zeta_m(\Y). \label{dzeta}
\end{eqnarray}
On the one hand, we take the derivative of Eq.(\ref{dzeta}) 
with respect to $q_\mu$ and use the Ward
identities (and the fact that $\zeta_2(\Y)$ and $\zeta_m(\Y)$
do not depend on $q$) to obtain
\begin{eqnarray}
-\bar\varpsi^1(\Y;q;\mu,0) &=& -\varpsi^1(\Y;q;\mu,0) - \zeta_2(\Y) \gamma^\mu.
\label{zeta0}
\end{eqnarray}

On the other hand, we take the functional derivative of Eq.(\ref{dzeta})
at $A=0$ and we obtain
\begin{eqnarray}
\bar\varpsi^1(\Y;q;\mu,p) &=& \varpsi^1(\Y;q;\mu,p) + \zeta_2(\Y)\gamma^\mu
-\zeta'_2(\Y) (\gamma\cdot q -m)\\&&-\zeta'_m(\Y), \label{fini}
\end{eqnarray}
where $\zeta'_2(\Y)$ and $\zeta'_m(\Y)$ denote the derivative
of $\zeta_2(\Y)$ and $\zeta_m(\Y)$ with respect to $A_\mu(p)$
at $A=0$.  The term $\zeta_2(\Y)\gamma^\mu$ comes from the
functional derivative of Eq.(\ref{varpsi0}).

If we take the value $p=0$ in Eq.(\ref{fini}) and compare with
Eq.(\ref{zeta0}) we obtain $\zeta'_2(\Y)=0$ and $\zeta'_m(\Y)=0$.
Further differentiation shows that $\zeta^{(n)}_2(\Y)=0$ and 
$\zeta^{(n)}_m(\Y)=0$. We can apply this proof to any tree $t$,
once the subdivergences have been subtracted. A similar proof
can be given for $\zeta_3(t)$, using Fury's theorem instead
of Ward identities. This proof assumes that the renormalization
conditions do not depend on $A$ (e.g. minimal subtraction).

More physically, $S(x,y;A)$ has the same singular
structure at $x=y$ as $S^0(x,y)$, except for logarithmic
terms that are integrable. Thus, the renormalization constants
are determined by $S^0(x,y)$.

This method of functional derivatives avoids the usual 
renormalization of vertex diagrams.
Renormalizing propagators is sufficient.

\section{The renormalized electron propagator}

In this section, we show that the recursive equation for the electron
propagator is
\begin{eqnarray}
\bar\varphi^0(t;q)&=&\rho(t)\varphi^0(\|;q)-\zeta_m(t)\varphi^0(\|;q)^2
\nonumber\\&&\hspace*{-5mm}
-\varphi^0(\|;q)(\zeta_m\star\bar\varphi^0(q))(t)
\nonumber\\&&\hspace*{-15mm}
+i \varphi^0(\|;q) \int \frac{\dd^4p}{(2\pi)^4}\gamma^{\lambda}
  \bar\varphi^0_{\lambda\lambda'}(t_l;p)
  \bar\varphi^1(t_r;q-p;\lambda',p),\label{recelec}
\end{eqnarray}
where
$\rho(t)=-\zeta_2(t)-(\rho\star\zeta_2)(t)=\zeta_2\circ\Sstar(t)$,
starting at $\rho(\Y)=-\zeta_2(\Y)$.

It is natural to define a new quantity $\alpha(t;q)$ by
$\alpha(t;q)=0$ for $t=\|$ and, for $t=t_l\vee t_r$,
\begin{eqnarray*}
\alpha(t;q)&=& ie^2 S^0(q)
  \int\frac{\dd^4p}{(2\pi)^4}
  \gamma^\lambda {\bar \varphi}_{\lambda\lambda'}(t_l;p)
  \bar\varphi^1(t_r;q-p;\lambda',p).\label{defalpha}
\end{eqnarray*}
Then we consider Eq.(\ref{eqbaseelectron}) and
in the integral over $p$, we expand
the photon propagator over trees $t_l$
using Eq.(\ref{renD}) and the electron
propagator over trees $t_r$ 
using Eq.(\ref{renS}).
We recognize a sum of $\alpha(t_l\vee t_r)$
and the integral becomes
\begin{eqnarray*}
 ie^2 S^0(q)
  \int\frac{\dd^4p}{(2\pi)^4}
  \gamma^\lambda {\bar D}_{\lambda\lambda'}(p)
  \frac{\delta {\bar S}(q-p)}{e\delta A_{\lambda'}(p)}  &=&
\sum_t \alpha(t;q).
\end{eqnarray*}

In the other terms of Eq.(\ref{eqbaseelectron}),
we expand $\bar S(q)$
and renormalization constants over trees using
Eqs.(\ref{renS}), (\ref{treeZ2}), (\ref{treeZ3}) and
(\ref{treeZm}). We replace products by convolutions 
$\nablaP$ according to Eq.(\ref{mulnablaP}) in all expressions
except the integral and we obtain.
\begin{eqnarray*}
\sum_t e^{|t|} (\bar\varphi^0(q)\nablaP\zeta_2)(t) & = &
\varphi^0(\|;q)
+ \sum_t e^{|t|} (\alpha(q)\nablaP\zeta_2)(t)
\\&&\hspace*{-10mm}-\sum_t e^{|t|} \varphi^0(\|;q)
  (\zeta_m\nablaP\bar\varphi^0(q)\nablaP\zeta_2)(t).
\end{eqnarray*}
The bold step is now to identify the terms corresponding
to a given tree $t$. This yields
\begin{eqnarray}
(\bar\varphi^0(q)\nablaP\zeta_2)(t) & = &
\varphi^0(\|;q)\epsilon(t)
+ (\alpha(q)\nablaP\zeta_2)(t)\nonumber\\&&
-\varphi^0(\|;q)
  (\zeta_m\nablaP\bar\varphi^0(q)\nablaP\zeta_2)(t),\label{ale}
\end{eqnarray}
where $\epsilon(t)=1$ if $t=\|$ and $\epsilon(t)=0$ 
otherwise.

To simplify this expression, we follow Kreimer
\cite{Kreimer} and compute
$(\bar\varphi^0(q)\nablaP\zeta_2\nablaP\zeta_2\circ\Sstar)(t)$.
The basic property of the antipode is
$Id\nablaP\Sstar=\|\epsilon$, therefore
\begin{eqnarray}
\zeta_2\nablaP\zeta_2\circ\Sstar=
  \zeta_2(Id\nablaP\Sstar)=\zeta_2\epsilon=\epsilon . \label{zeta21}
\end{eqnarray}

The associativity of $\nablaP$ is here crucial.
On the one hand, 
$\big(\bar\varphi^0(q)\nablaP
(\zeta_2\nablaP\zeta_2\circ\Sstar)\big)(t)
=\bar\varphi^0(q)(t)$ according to (\ref{zeta21}).
On the other hand, we calculate
$\big((\bar\varphi^0(q)\nablaP\zeta_2)\nablaP\zeta_2\circ\Sstar)\big)$,
where we replace $\bar\varphi^0(q)\nablaP\zeta_2$
by the right-hand side of Eq.(\ref{ale}).
Equation (\ref{zeta21}) gives us 
\begin{eqnarray*}
\big((\bar\varphi^0(q)\nablaP\zeta_2)\nablaP\zeta_2\circ\Sstar)\big) &=&
\zeta_2\circ \Sstar(t)\varphi^0(\|;q)\\&&\hspace{-10mm}-
\varphi^0(\|;q)(\zeta_m\nablaP\bar\varphi^0(q))(t)+\alpha(t).
\end{eqnarray*}
From the associativity of $\nablaP$ and the definition
of $\alpha(t;q)$ 
we obtain our final recursive equation
(\ref{recelec}) for the electron propagator.

The recursive equation is completed by the
equation for the higher components of $\bar\varphi^0(t;q)$.
If we take the functional derivative of Eq.(\ref{recelec}),
and make the same simplification as in Ref.\cite{BrouderEPJC2}),
we obtain
\begin{eqnarray}
\bar\varphi^n(t;q;\{\lambda,p\}_{1,n})&=&
\varphi^0(\|;q)\gamma^{\lambda_1}\bar
  \varphi^{n-1}(t;q;\{\lambda,p\}_{2,n})
\nonumber\\&&\hspace*{-20mm}
-\rho(t)\varphi^0(\|;q)\delta_{n,0}
-\zeta_m(t)\varphi^0(\|;q)\bar\varphi^n(\|;q;\{\lambda,p\}_{1,n})
\nonumber\\&&\hspace*{-20mm}
-\varphi^0(\|;q)(\zeta_m\star\bar\varphi^n(q;\{\lambda,p\}_{1,n}))(t)
\nonumber\\&&\hspace*{-20mm}
+i \varphi^0(\|;q) \sum_{k=0}^n\int \frac{\dd^4p}{(2\pi)^4}\gamma^{\lambda}
  \bar\varphi^k_{\lambda\lambda'}(t_l;p;\{\lambda,p\}_{1,k})
\nonumber\\&&\hspace*{-20mm}
  \times\bar\varphi^{n-k+1}_\Sigma(t_r;q-p;\lambda',p,\{\lambda,p\}_{k+1,n}).
\label{recelecn}
\end{eqnarray}

Another useful formula can be obtained by defining 
\begin{eqnarray}
\bar{f}^0(t;q) &=& (\bar\varphi^0(q)\nablaP\zeta_2)(t)\\&=&
\bar\varphi^0(t;q)+
(\bar\varphi^0(q)\star\zeta_2)(t)+
\varphi^0(\|;q)\zeta_2(t),\nonumber\\
\bar{f}^1(t;q;\lambda',p) &=&
(\bar\varphi^1(q;\lambda',p)\nablaP\zeta_2)(t)\label{deff}.
\end{eqnarray}

With this notation Eq.(\ref{ale}) is rewritten
\begin{eqnarray}
\bar{f}^0(t;q)&=&-\zeta_m(t)\varphi^0(\|;q)^2
-\varphi^0(\|;q)(\zeta_m\star\bar{f}^0(q))(t)\nonumber\\&&\hspace*{-16mm}
+i \varphi^0(\|;q)\int \frac{\dd^4p}{(2\pi)^4}\gamma^{\lambda}
  \bar\varphi^0_{\lambda\lambda'}(t_l;p)
  \bar{f}^1(t_r;q-p;\lambda',p). \label{eqbarf}
\end{eqnarray}

The higher components $\bar{f}^n(t;q)$ are obtained by functional
derivative of (\ref{eqbarf}), as explained in section \ref{compon}.
The recursive equation for $\bar{f}^n(t;q)$ is the same
as Eq.(\ref{recelecn}), where $\varphi$ is replaced by $f$ and
the term $\rho(t)\varphi^0(\|;q)\delta_{n,0}$ is suppressed.

\section{The renormalized photon propagator}
Bogoliubov and Shirkov \cite{Bogoliubov} have shown that
the renormalization of
two Feynman diagrams linked by a single photon (or electron)
line is obtained by an independent renormalization
of each of the two subgraphs. In our language,
this means that the renormalized form of (\ref{phiphi2})
is
\begin{eqnarray}
\bar\varphi^0_{\mu\nu}(t_l\vee t_r;q)&=&
\bar\varphi^0_{\mu\lambda}(\|\vee t_r;q)
\varpsi^{0\lambda\lambda'}(\|;q)
\bar\varphi^{0}_{\lambda'\nu}(t_l;q). \label{phiphir}
\end{eqnarray}

Therefore, all trees for the photon propagator 
can be renormalized once we have renormalized the
special trees $\|\vee t_r$. Now we show that
the recursive equation for the renormalized photon term
$\bar\varphi^0_{\mu\lambda}(\|\vee t_r;q)$ is
\begin{eqnarray}
\bar\varphi^0_{\mu\nu}(\|\vee t;q)&=&-\zeta_3(\|\vee t)
 \varphi^T_{\mu\nu}(\|;q)
  \nonumber\\&&\hspace{-23mm}-
 i \varphi^T_{\mu\lambda}(\|;q) \int \frac{\dd^4p}{(2\pi)^4}
  \mathrm{tr}\big[\gamma^\lambda
  \bar{f}^1(t;p;\lambda',-q)
  \big] \varphi^T_{\lambda'\nu}(\|;q), \label{barphif}
\end{eqnarray}
where $\bar{f}^1$ was defined in the previous section.

To prove this, we start from several remarks:
we have $\bar\varpsi^0_{\mu\nu}(t_l\vee t_r)=0$ is $t_l\not= \|$
and 
\begin{eqnarray*}
\bar\varpsi^0_{\mu\nu}(\|\vee t_r;q)&=&
-\varpsi^0_{\mu\lambda}(\|;q)
\bar\varphi^{0\lambda\lambda'}(\|\vee t_r;q)
\varpsi^0_{\lambda'\nu}(\|;q).
\end{eqnarray*}

Because of this close analogy between photon propagator
and vacuum polarization, we shall rewrite (\ref{eqbasephoton})
as
\begin{eqnarray*}
\sum_t e^{2|t|}{\bar\varphi}^0_{\mu\nu}(\|\vee t)&= & 
-\varphi^T_{\mu\lambda}(\|;q){\bar\Pi}^{\lambda\lambda'}(q) 
\varphi^T_{\lambda'\nu}(\|;q)
\\&= & 
-(Z_3-1) \varphi^{T}_{\lambda\mu}(\|;q)
\\&&\hspace*{-23mm}
- Z_2 ie^2 \varphi^T_{\mu\lambda}(\|;q)\int \frac{\dd^4p}{(2\pi)^4}
  \mathrm{tr}\big[\gamma^\lambda
  \frac{\delta {\bar S}(p)}{e \delta A_{\lambda'}(-q)}\big]
  \varphi^T_{\lambda'\nu}(\|;q).
\end{eqnarray*}

We rewrite this expression
to isolate the root components:
\begin{eqnarray*}
\sum_t {\bar\varphi}^0_{\mu\nu}(\|\vee t)&= & 
-(Z_3-1) \varphi^{T}_{\lambda\mu}(\|;q)
\\&&\hspace*{-20mm}
- ie^2 \varphi^T_{\mu\lambda}(\|;q)\int \frac{\dd^4p}{(2\pi)^4}
  \mathrm{tr}\big[\gamma^\lambda
  \frac{\delta {\bar S}(p)}{e \delta A_{\lambda'}(-q)}\big]
  \varphi^T_{\lambda'\nu}(\|;q)
\\&&\hspace*{-20mm}
+ ie^2 \varphi^T_{\mu\lambda}(\|;q)\int \frac{\dd^4p}{(2\pi)^4}
  \mathrm{tr}\big[\gamma^\lambda
  \frac{\delta {\bar S}(p)}{e \delta A_{\lambda'}(-q)}\big](Z_2-1)\\&&
  \times\varphi^T_{\lambda'\nu}(\|;q).
\end{eqnarray*}

We expand all quantities over trees, using
\begin{eqnarray*}
\frac{\delta {\bar S}(p)}{e \delta A_{\lambda'}(-q)}&=&
\sum_t e^{2|t|}{\bar\varphi}^1(t;p;\lambda',-q),
\end{eqnarray*}
and we multiply through the pruning operator. Then we identify
the terms corresponding to the same tree and we obtain
\begin{eqnarray*}
\bar\varphi^0_{\mu\nu}(\|\vee t;q)&=&-\zeta_3(\|\vee t) 
  \varphi^T_{\mu\nu}(\|;q)+
  \zeta_2(t)\varphi^0_{\mu\nu}(\Y;q)\nonumber\\&&\hspace{-21mm}-
 i \varphi^T_{\mu\lambda}(\|;q) \int \frac{\dd^4p}{(2\pi)^4}
  \mathrm{tr}\big[\gamma^\lambda
  \bar\varphi^{1}(t;p;\lambda',-q)
  \big] \varphi^T_{\lambda'\nu}(\|;q)\nonumber\\&&\hspace{-21mm}-
 i \varphi^T_{\mu\lambda}(\|;q)\int \frac{\dd^4p}{(2\pi)^4}
  \mathrm{tr}\big[\gamma^\lambda
  (\bar\varphi^{1}\star\zeta_2)(t;p;\lambda',-q)
  \big] \varphi^T_{\lambda'\nu}(\|;q).
\end{eqnarray*}

From the definition (\ref{deff}) for $\bar f^1$,
we can rewrite this expression as
our recursive equation (\ref{barphif}).

The higher components are obtained very simply by
taking the functional derivative of  Eq.(\ref{barphif}).
Since $\varphi^T_{\mu\nu}(\|;q)$ is independent of the
external field, we obtain
\begin{eqnarray*}
\bar\varphi^n_{\mu\nu}(\|\vee t;q;\{\lambda,p\}_{1,n})&=&
-i \varphi^T_{\mu\lambda}(\|;q) 
  \nonumber\\&&\hspace{-28mm}
\times
 \int \frac{\dd^4p}{(2\pi)^4}
  \mathrm{tr}\big[\gamma^\lambda
  \bar{f}^{n+1}_\Sigma(t;p;\lambda',-q,\{\lambda,p\}_{1,n})
  \big] \varphi^T_{\lambda'\nu}(\|;q).
\end{eqnarray*}

For the other trees, we use
\begin{eqnarray*}
\bar\varphi^n_{\mu\nu}(t_l\vee t_r;q;\{\lambda,p\}_{1,n})&=&
\sum_{k=0}^n\bar\varphi^k_{\mu\lambda}(\|\vee t_r;q;\{\lambda,p\}_{1,k})
\\&&\hspace*{-10mm}\times
\varpsi^{0\lambda\lambda'}(\|;q)
\bar\varphi^{n-k}_{\lambda'\nu}(t_l;q;\{\lambda,p\}_{k+1,n}).
\end{eqnarray*}

\subsection{Properties of renormalized photon propagator}

From (\ref{eqbarf}) and (\ref{barphif}) we can deduce that
the renormalized photon propagator does not depend on
any $\zeta_2(t)$. 
In fact, we shall prove that
$\bar{f}^0(t;q)$, $\bar{f}^1(t;q;\lambda,p)$
and $\bar{\varphi}^0_{\mu\nu}(t;q)$ do not depend on any 
$\zeta_2(t')$.
To do this, we reintroduce a non-zero external field $A$.
The property is clearly true for $t=\|$. If it is true for
all trees with $|t|<N$, let us take a tree with $|t|=N$.
Because of (\ref{eqbarf}), $\bar{f}^0(t;q)$ does not
depend on any $\zeta_2(t')$. Since $\bar{f}^1(t;q;\lambda,p)$
is obtained by a functional derivative of $\bar{f}^0(t;q)$
with respect to $eA$, it does not depend on any $\zeta_2(t')$
either ($eA$ does not depend on any $\zeta_2(t')$).
If $t=\|\vee t_r$, because of (\ref{barphif}), 
$\bar\varphi^0_{\mu\nu}(\|\vee t_r;q)$ does not depend on any
$\zeta_2(t')$ since none of the terms on the right hand side do.
Finally, if $t$ is not of the form
$t=\|\vee t_r$, it is of the form $t=t_l \vee t_r$, and
$\bar\varphi^0_{\mu\nu}(t_l \vee t_r;q)$ is obtained from 
$\bar\varphi^0_{\mu\nu}(\| \vee t_r;q)$ and $\bar\varphi^0_{\mu\nu}(t_l;q)$,
which do not depend on any $\zeta_2(t')$.

With the same reasoning, we see that 
$\bar{f}^0(t;q)$ and $\bar\varphi^0_{\mu\nu}(t;q)$
are independent of the gauge parameter $\xi$ for
$t\not=\|$.

\section{Electron self-energy}

To calculate the electron self-energy, we start from (\ref{recelec})
that we rewrite
\begin{eqnarray*}
\bar\varphi^0(t;q) &=& \rho(t) \varphi^0(\|;q) 
 -\zeta_m(t)\varphi^0(\|;q)^2+\alpha(t;q)\\&&
 -\varphi^0(\|;q)(\zeta_m\star\bar\varphi^0(q))(t).
\end{eqnarray*}

The self-energy is obtained by introducing the last equation
into (\ref{inversepsi}). This gives us
\begin{eqnarray*}
\bar\varpsi^0(t;q) &=& -\rho(t) \varpsi^0(\|;q) 
 -\zeta_m(t)+ \varpsi^0(\|;q)\alpha(t;q)\varpsi^0(\|;q)
\\&&\hspace*{-15mm}
 -(\rho\star\bar\varpsi^0(q))(t)
 +(\zeta_m\star\bar\varphi^0(q))(t)\varpsi^0(\|;q)
\\&&\hspace*{-15mm}
 +\varphi^0(\|;q)(\zeta_m\star\bar\varpsi^0(q))(t) 
 +(\zeta_m\star\bar\varphi^0(q)\star\bar\varpsi^0(q))(t)
\\&&\hspace*{-15mm}
 +\varpsi^0(\|;q)(\alpha(q)\star\bar\varpsi^0(q))(t).
\end{eqnarray*}
If we factorize $\zeta_m$ and use (\ref{inversephi}), the
expression reduces to
\begin{eqnarray*}
\bar\varpsi^0(t;q) &=& -\rho(t) \varpsi^0(\|;q) 
 -\zeta_m(t)+ \varpsi^0(\|;q)\alpha(t;q)\varpsi^0(\|;q)\\&&
 -(\rho\star\bar\varpsi^0(q))(t)
 +\varpsi^0(\|;q)(\alpha(q)\star\bar\varpsi^0(q))(t).
\end{eqnarray*}

From the definition of $\alpha(t;q)$ and the result of 
section \ref{deuxg}, we obtain 
\begin{eqnarray*}
\bar\varpsi^0(t;q) &=& -\rho(t) \varpsi^0(\|;q) 
 -\zeta_m(t) -(\rho\star\bar\varpsi^0(q))(t)\\&&
 + i \int \frac{\dd^4p}{(2\pi)^4}\gamma^{\lambda}
  \bar\varphi^0_{\lambda\lambda'}(t_l;p)
  \bar{g}(t_r;q-p;\lambda',p), \label{intpsi}
\end{eqnarray*}
where
\begin{eqnarray}
\bar{g}(t_r;q-p;\lambda',p) &=&
-(\bar\varphi^1(q-p;\lambda',p)\nablaP\bar\varpsi^0(q))(t_r)\\
&=&
(\bar\varphi^0(q)\nabla\bar\varpsi^1(q-p;\lambda',p))(t_r). \label{defg}
\end{eqnarray}

It can be shown recursively that the only term proportional to
$\varpsi^0(\|;q)$ in $-\rho(t)\varpsi^0(\|;q)-(\rho\star\bar\varpsi^0(q))(t)$
is $\zeta_2(t)\varpsi^0(\|;q)$.

\section{Renormalization conditions}

Renormalization conditions are the conditions used to
give a unique value to the renormalized quantities. As noticed
by Itzykson and Zuber \cite{Itzykson}, there is some freedom
in the choice of these conditions. However, it is necessary
that the conditions are satisfied for the root. In other
words, the renormalization conditions can be non-zero only
for the root.
The physical meaning of these renormalization conditions
can be found in any textbook on quantum field theory.
In particular, the relation between the renormalization
conditions and the introduction of experimental quantities
into the theory is discussed at length in Refs.
\cite{Itzykson,Collins}.

For instance, mass shell renormalization conditions 
for QED \cite{Itzykson} p.413 could be translated into
(for $t\not=\|$)
\begin{eqnarray*}
\bar\varpsi^0(t;q)|_{\gamma^\mu q_\mu=m} &=& 0,\\
\bar\varpsi^1(t;q;\lambda,0)|_{\gamma^\mu q_\mu=m} &=& 0,\\
\bar\omega(t;q)|_{q=0} &=& 0.
\end{eqnarray*}

The strange prescription $\gamma^\mu q_\mu=m$ means that
the quantities $\bar\varpsi^0(t;q)$ and
$\bar\varpsi^1(t;q;\lambda,0)$ must be multiplied
by $u(q)$ on the right where $u(q)$ is a solution
of the Dirac equation $(\gamma^\mu q_\mu-m)u(q)=0$.
This amounts to replacing $q^2$ by $m^2$ and
$\gamma^\mu q_\mu$ by $m$ in the analytic expressions
for $\bar\varpsi^0(t;q)$ and
$\bar\varpsi^1(t;q;\lambda,0)$, when they are available.

For the photon, we
define  $\bar\omega(q^2)$ by
\begin{eqnarray}
{[{\bar D}^{-1}]}_{\lambda\mu}(q) & = & (q_\lambda q_\mu -q^2 g_{\lambda\mu})
(1+{\bar\omega}(q^2))-\xi q_\lambda q_\mu .\label{Domega}
\end{eqnarray}
This $\bar\omega(q^2)$ is the same as $\bar\omega_R(q^2)$
in Itzykson
and Zuber \cite{Itzykson}.

For the photon, we know that $\psi^0_{\mu\nu}(t;q)$
is transverse. Therefore, we can define
$\bar\omega(\|\vee t;q)$ by
\begin{eqnarray*}
\psi^0_{\mu\nu}(\|\vee t;q)=\psi^0_{\mu\nu}(\|;q)
\bar\omega(\|\vee t;q). 
\end{eqnarray*}
Notice that $\sum_t \bar\omega(\|\vee t;q)=\bar\omega(q)$,
where $\bar\omega(q)$ was defined in (\ref{Domega}).
The renormalization
condition on $\bar\omega(\|\vee t;q)$ replaces
a more complicated renormalization condition on
$\psi^0_{\mu\nu}(t;q)$ (involving the
third derivative of $\psi^0_{\mu\nu}(t;q)$
with respect to $q$).

To show how this works in practice, we use the identity
\begin{eqnarray*}
\bar\varphi^0_{\lambda\mu}(\|\vee t;q)=
-\bar\omega(\|\vee t;q^2) \bar\varphi^T_{\lambda\mu}(\|;q)
\end{eqnarray*}
Now we assume that we have calculated the integral
\begin{eqnarray*}
h(q)&=&
-\frac{i}{q^2} g^{\lambda\lambda'} \int \frac{\dd^4p}{(2\pi)^4}
  \mathrm{tr}\big[\gamma^\lambda
  \bar{f}^1(t;p;\lambda',-q)
  \big].
\end{eqnarray*}
This integral is logarithmically divergent, and we must 
determine how to remove its divergence.
We start from Eq.(\ref{barphif}), that we multiply by
$g^{\mu\nu}$ and sum over $\mu$ and $\nu$. Using the
fact that the integral over $p$ is transverse, we obtain
\begin{eqnarray*}
3\bar\omega(\|\vee t;q^2) &=& 3\zeta_3(\|\vee t) + h(q).
\end{eqnarray*}
Since the renormalization condition is $\bar\omega(\|\vee t;0)=0$,
we just take $\zeta_3(\|\vee t)=-h(0)/3$. This satisfies the
renormalization condition and this determines $\zeta_3(\|\vee t)$.
For the electron propagator, the renormalization conditions
determine $\zeta_2(t)$ and $\zeta_m(t)$.

In practice, other renormalization conditions are
used, such as minimal subtraction. Once quantities
have been made convergent by a renormalization condition,
it is always possible to translate the results
into different renormalization conditions,
using the Hopf structure of renormalization 
\cite{Kreimer}.
The fact
that the renormalization conditions can be composed
and that the result does not depend on which condition
is used first is a direct consequence of the
coassociativity of the Hopf algebra of renormalization
\cite{Kreimer}.

\section{Massive quenched QED}

Quenched QED is QED without vacuum insertion graphs.
Quenched QED has been recently advocated and discussed by
Broadhurst and coll. \cite{BroadhurstQI,BroadhurstQII}.

The Feynman diagrams describing the electron propagator
of quenched QED have no fermion loop.
Within the present approach, all Feynman diagrams  of
a given order for quenched QED are summed into one tree $\varphi^0(\Y^n)$.
The trees $\Y^n$ are defined recursively by
$\Y^0=\|$, $\Y^n=\|\vee\Y^{n-1}$. In words,
they are the trees without left branches.
The Feynman diagrams corresponding to
$\varphi^0(\Y^n)$ for $n\le 3$
are given in appendix 3.

The bare and renormalized electron propagators
for quenched QED are written
\begin{eqnarray*}
S^Q(q) &=& \sum_{n=0}^\infty e_0^n \varphi^0(\Y^n;q)
\quad\mathrm{and}\quad
{\bar{S}}^Q(q) = \sum_{n=0}^\infty e^n \bar\varphi^0(\Y^n;q).
\end{eqnarray*}

The Schwinger equation for the renormalized electron
propagator of quenched QED is
\begin{eqnarray}
{\bar{S}}^Q(q) &=& S^0(q)-(Z_2-1){\bar{S}}^Q(q)-
  \delta m Z_2 S^0(q){\bar{S}}^Q(q)
 \nonumber\\&&\hspace*{-6mm}+
 ie^2 Z_2 S^0(q)
  \int\frac{\dd^4p}{(2\pi)^4}
  \gamma^\lambda D^0_{\lambda\lambda'}(p)
  \frac{\delta {\bar{S}}^Q(q-p)}{e\delta A_{\lambda'}(p)}.
  \label{QSchwS}
\end{eqnarray}

The relation we want to show is

\begin{eqnarray}
\bar\varphi^0(\Y^n;q)&=&\varphi^0(\Y^n;q)+
  \sum_{a=0}^{n-1}\sum_{k=0}^{n-a}
  \frac{1}{k!}
  \alpha_k(\Y^{n-a})\frac{\partial^k\varphi^0(\Y^a;q)}
  {\partial m^k}\nonumber\\
&=&\sum_{k=0}^{n}
  \frac{1}{k!} \big(\alpha_k\nabla\frac{\partial^k\varphi^0(q)}
  {\partial m^k}\big) (\Y^n), \label{Qelectron}
\end{eqnarray}
where
\begin{eqnarray*}
\alpha_0 &=& \rho = -\zeta_2-\rho\star\zeta_2,\\
\alpha_1 &=& -\zeta_m -\zeta_m\star\alpha_0,\\
\alpha_k &=& -\zeta_m\star\alpha_{k-1}\quad \mathrm{if}\quad k\ge 2.
\end{eqnarray*}
Notice that $\alpha_k(\Y^n)=0$ if $k>n$, because $\Y^n$ cannot
be split into more than $n$ part by the pruning operator $P$.

To prove this, we start from the Schwinger equation (\ref{QSchwS})
and transform it into a recurrence relation:
\begin{eqnarray}
\bar\varphi^0(t;q)&=&\rho(t)\varphi^0(\|;q)-\zeta_m(t)\varphi^0(\|;q)^2
\nonumber\\&&\hspace*{-5mm}
-\varphi^0(\|;q)(\zeta_m\star\bar\varphi^0(q))(t)
\nonumber\\&&\hspace*{-15mm}
+i S^0(q) \int \frac{\dd^4p}{(2\pi)^4}\gamma^{\lambda}
  D^0_{\lambda\lambda'}(p)
  \bar\varphi^1(t_r;q-p;\lambda',p),\label{Qrecelec}
\end{eqnarray}
where
$\rho(t)=-\zeta_2(t)-(\rho\star\zeta_2)(t)=\zeta_2\circ\Sstar(t)$,
starting at $\rho(\Y)=-\zeta_2(\Y)$.

Using the definitions given in section \ref{compon},
we can show that
\begin{eqnarray*}
\frac{\delta}{e\partial A_\lambda(p)} 
\frac{\partial}{\partial m} \varphi^0(\Y^n;q)
 &=& 
\frac{\partial}{\partial m}
\frac{\delta}{e\partial A_\lambda(p)} 
\varphi^0(\Y^n;q),
\end{eqnarray*}
and that
\begin{eqnarray*}
\frac{\partial^k}{\partial m^k} \varphi^0(\Y^n;q) &=&
k\varphi^0(\|;q) \frac{\partial^{k-1}}{\partial m^{k-1}} \varphi^0(\Y^n;q)
\\&&\hspace*{-20mm}
+i\varphi^0(\|;q) 
 \int \frac{\dd^4p}{(2\pi)^4}\gamma^{\lambda}
  D^0_{\lambda\lambda'}(p)
  \frac{\partial^k}{\partial m^k}\varphi^1(t_r;q-p;\lambda',p).
\end{eqnarray*}
From these identities, Eq.(\ref{Qelectron}) follows easily
by recursion.

The Feynman diagrams describing the photon propagator of 
quenched QED have a single electron loop,
as can be seen in the diagrams for
$\varphi^0_{\lambda\mu}(\Y^n)$ in appendix 3.
The bare and renormalized photon propagators are written
\begin{eqnarray*}
D^Q_{\lambda\mu}(q) &=& \sum_{n=0}^\infty e_0^n 
  \varphi^0_{\lambda\mu}(\Y^n;q),\\
{\bar{D}}^Q_{\lambda\mu}(q) &=& \sum_{n=0}^\infty e^n 
  \bar\varphi^0_{\lambda\mu}(\Y^n;q).
\end{eqnarray*}

The relation we want to show is 
\begin{eqnarray}
\bar\varphi^0_{\lambda\mu}(\Y^n;q)&=&\varphi^0_{\lambda\mu}(\Y^n;q)+
  \sum_{k=0}^{n-1}\sum_{a=1}^{n-k}
  \frac{1}{k!}\beta_k(\Y^{n-a}) 
  \frac{\partial^k\varphi^0_{\lambda\mu}(\Y^a;q)}
  {\partial m^k}\nonumber\\&&
  -\zeta_3(\Y^n) \varphi^T_{\lambda\mu}(\|;q),
\label{Qphoton}
\end{eqnarray}
where the scalars $\beta_k$ are defined recursively
\begin{eqnarray*}
\beta_1 &=& -\zeta_m,\\
\beta_k &=& -\zeta_m\star\beta_{k-1}\quad k\ge 2,
\end{eqnarray*}
with $\beta_k(\Y^n)=0$ if $k>n$.
This can be written more compactly:
\begin{eqnarray*}
\bar\varphi^0_{\lambda\mu}(\Y^n;q)
&=&\exp\big[-\zeta_m\frac{\partial}{\partial m}\star\big]
\varphi^0_{\lambda\mu}(\Y^n;q)
-\zeta_3(\Y^n) \varphi^T_{\lambda\mu}(\|;q).
\end{eqnarray*}

To prove Eq.(\ref{Qphoton}), we first use our previous
result Eq.(\ref{Qelectron}) to show that
\begin{eqnarray*}
\bar{f}^0(\Y^n;q)&=&\varphi^0(\Y^n;q)+
  \sum_{k=1}^{n}\sum_{a=0}^{n-k}
  \frac{1}{k!} \beta_k(\Y^{n-a})\frac{\partial^k\varphi^0(\Y^a;q)}
  {\partial m^k}.
\end{eqnarray*}

Then we introduce this expression into Eq.(\ref{barphif}),
and the result follows from the relation
\begin{eqnarray}
\frac{\partial^k}{\partial m^k}\varphi^0_{\mu\nu}(\Y^{n+1};q)&=&
 -i \varphi^T_{\mu\lambda}(\|;q) 
  \nonumber\\&&\hspace{-30mm}\times
  \int \frac{\dd^4p}{(2\pi)^4}
  \frac{\partial^k}{\partial m^k}\mathrm{tr}\big[\gamma^\lambda
  \bar{f}^1(\Y^n;p;\lambda',-q)
  \big] \varphi^T_{\lambda'\nu}(\|;q).
\end{eqnarray}

\section{Hopf algebra for massless QED}

In this section, we determine a coproduct from the 
recursive equations (\ref{recelec}),  (\ref{eqbarf})
and (\ref{barphif}) and the product law (\ref{phiphir})
for massless QED. The case of massless QED is much
simpler because the mass is not renormalized. Thus,
for all trees $t$, $\zeta_m(t)=0$.

This coproduct determines the renormalized propagators
as a function of the unrenormalized ones. In this section,
it will be useful to dinstinguish the electron and
photon trees by the color of the root. A tree with
a black root is written $t^\bullet$ and represents an
electron propagator, a tree with a white root is written
$t^\circ$ and represents a photon propagator.
In a tree $t_l\vee t_r$, $t_l$ is white and $t_r$ is black.
There are now two graftings operators $\veeb$ and $\veen$,
so that $t_l\veen t_r$ is a black tree and $t_l\veeb t_r$
a white one.

Using a variation of Sweedler's notation, we write
\begin{eqnarray}
\Delta t^\circ &=& \sum_{\Delta t^\circ} 
   t^{\circ}_{(1)}\otimes t^{\circ}_{(2)},\label{Deltab}\\
\Delta t^\bullet &=& \sum_{\Delta t^\bullet} 
  t^{\circ}_{(1)}t^{\bullet}_{(1)}\otimes t^{\bullet}_{(2)},\label{Deltan}\\
F(t^\bullet)&=& \sum_{F(t^\bullet)} t^{\circ}_{(1)}\otimes t^{\bullet}_{(2)}.
\label{Deltaf}
\end{eqnarray}
These equations mean that the coproduct of $t^\circ$
generates a sum of tensor products with one white tree on
the left and one white tree on the right,
the coproduct of $t^\bullet$
generates a sum of tensor products with one black tree and
one white tree on
the left and one black tree on the right, finally the
coproduct $F(t)$ generates a sum of tensor products with 
one white tree on
the left and one black tree on the right.
These trees can eventually be the root, which is the unit element
of the algebra (the root is neither white nor black, or both, as you wish).

To avoid products of white trees in Eqs.(\ref{Deltab}), 
(\ref{Deltan}) and (\ref{Deltaf}), we took advantage of
the fact that, according to Eq.(\ref{phiphi2}),
the $\varphi_{\mu\nu}$ of a white tree $t_l\veeb t_r$
can be written as a product of $\varphi_{\mu\nu}(\|\veeb t_r)$
by $\varphi_{\mu\nu}(t_l)$. From Eq.(\ref{phiphir}), we
also know that this property is compatible with renormalization.
Therefore, we 
translate this property into an inner product over white trees.
The product of two white trees $s^\circ / t^\circ$
(read ``$s$ over $t$'') is defined recursively by
$s/\|=s$ and $s/(t_l\veeb t_r)=(s/t_l)\veeb t_r$.
In particular $t_l/(\|\veeb t_r)=t_l\veeb t_r$,
which is what we need.
Surprisingly, this product has been used
independently by Loday and Ronco in a completely 
different context \cite{LodayRonco2}.

The coproduct $\Delta$ acting on white and black trees is defined by the
recursive equations
\begin{eqnarray}
\Delta (\|\veeb t) &=& (\|\veeb t)\otimes \|
+ \sum_{F(t)} t^{\circ}_{(1)}\otimes (\|\veeb t^{\bullet}_{(2)}),
\label{Deltarecb}\\
\Delta (t_l\veen t_r) &=& (t_l\veen t_r)\otimes \|
+ \sum_{\Delta t_l,\Delta t_r} 
  (t^\circ_{l(1)}/t^\circ_{r(1)}) t^\bullet_{r(1)} \nonumber\\&&
  \otimes (t_{l(2)}^\circ\veen t_{r(2)}^\bullet)
  ,\label{Deltarecn}\\
F(t_l\veen t_r)&=& \sum_{\Delta t_l,F(t_r)} 
  (t^\circ_{l(1)}/ t^\circ_{r(1)}) \otimes 
(t_{l(2)}^\circ\veen t_{r(2)}^\bullet),
\label{Deltarecf}
\end{eqnarray}
with the initial values $\Delta \|=\|\otimes\|$ and 
$F(\|)=\|\otimes\|$, and with the compatibility of 
the ``over'' product with renormalization:
$\Delta (s^\circ/t^\circ)=\Delta t^\circ \Delta s^\circ.$
In particular,
\begin{eqnarray*}
\Delta t_l\veeb t_r=\Delta (\|\veeb t_r) \Delta t_l. 
\end{eqnarray*}

With this notation, we can now write the coproduct of
a general white tree
\begin{eqnarray}
\Delta (t_l \veeb t_r) &=& \sum_{\Delta t_l }(t^\circ_{l(1)}
 \veeb t_r)\otimes t^\circ_{l(2)}\nonumber\\&&
+ \sum_{\Delta t_l,F(t_r)} (t^\circ_{l(1)}/t^\circ_{r(1)})
  \otimes (t_{r(2)}^\circ \veeb t^{\bullet}_{(2)}).
\label{Deltagenb}
\end{eqnarray}

These preliminaries enable us to write the relation between
renormalized and unrenormalized propagators as
\begin{eqnarray}
\bar\varphi^0_{\mu\nu}(t^\circ;q) &=& \sum_{\Delta t^\circ} 
   \zeta(t^{\circ}_{(1)}) \varphi^0_{\mu\nu}(t_{(2)}^\circ;q),\label{Deltabres}\\
\bar\varphi^0(t^\bullet;q) &=& \sum_{\Delta t^\bullet} 
  \zeta(t^{\circ}_{(1)})\zeta(t^{\bullet}_{(1)})\varphi^0(t^{\bullet}_{(2)};q),
      \label{Deltanres}\\
\bar f(t;q)&=& \sum_{\Delta t^\bullet} \zeta(t^{\circ}_{(1)}) f^0(t^{\bullet}_{(2)}).
\label{Deltafres}
\end{eqnarray}

The general counterterm $\zeta$ is a scalar over black and
white trees defined by $\zeta(\|)=1$ and
\begin{eqnarray}
\zeta(t^\bullet)&=&\rho(t^\bullet),\label{zetarho}\\
\zeta(s^\circ / t^\circ)&=&\zeta(s^\circ)\zeta(t^\circ),\nonumber\\
\zeta(\|\veeb t)&=&-\zeta_3(\|\vee t).\nonumber
\end{eqnarray}
In particular $\zeta(t_l\veeb t_r)=-\zeta_3(\|\vee t_r)\zeta(t_l)$.
We recall that $\rho(t)=\zeta_2\circ \Sstar(t)$.
Equations (\ref{Deltabres}) and (\ref{Deltanres}) are given in
expanded form in appendix 2 for trees up to order 3.

We prove this recursively. From the list of appendix 2,
Eqs.(\ref{Deltabres}) and (\ref{Deltanres}) are satisfied
for all trees up to order 3. The same can be checked for
Eq.(\ref{Deltafres}). Assume that they are satisfied up
for trees with $2N-1$ vertices. Take a tree with
$2N+1$ vertices. Take first $\|\veeb t$, then use 
Eq.(\ref{Deltafres}) for $\bar f^1$ in Eq.(\ref{barphif}).
This yields
\begin{eqnarray*}
\bar\varphi^0_{\mu\nu}(\|\vee t;q)&=&-\zeta_3(\|\vee t)
 \varphi^T_{\mu\nu}(\|;q)\\&&
+\sum_{F(t)} \zeta(t_{(1)}) \varphi^0_{\mu\nu}(\|\vee t_{(2)}).
\end{eqnarray*}
This is Eq.(\ref{Deltabres}) for the coproduct defined
by Eq.(\ref{Deltarecb}).
If we take now $t_l\veeb t_r$, where $t_l$ and $t_r$ have
less than $2N+1$ vertices, we can expand 
$\bar\varphi^0_{\mu\nu}(\|\vee t_r)$
and $\bar\varphi^0_{\mu\nu}(t_l)$ over unrenormalized
terms. Then, using Eq.(\ref{phiphir}) we find
\begin{eqnarray*}
\bar\varphi^0_{\mu\nu}(t_l\vee t_r;q) &=& 
\Big(-(\zeta_3(\|\vee t_r)
 \varphi^T_{\mu\nu}(\|;q)
  \nonumber\\&&\hspace{-23mm}
+\sum_{F(t_r)} \zeta(t_{r(1)}) 
\varphi^0_{\mu\nu}(\|\vee t_{r(2)};q)\Big)
\sum_{\Delta t_l} \zeta(t_{l(1)}) 
\bar\varphi^0_{\mu\nu}(t_{l(2)};q)\\
&=& \sum_{\Delta t_l} \zeta(t_{l(1)}\veeb t_r)
  \bar\varphi^0_{\mu\nu}(t_{l(2)};q)
\\&+& \sum_{F(t_r) \Delta t_l} \zeta(t_{l(1)}/t_{r(1)})
\bar\varphi^0_{\mu\nu}(t_{l(2)}\vee t_{r(2)};q).
\end{eqnarray*}
This is Eq.(\ref{Deltabres}) with the coproduct defined
in Eq.(\ref{Deltagenb}) and the correct $\zeta$.

For the coproduct acting on electron trees, we
start from the recursive equation (\ref{recelec}) and
we use the expansion (\ref{Deltabres}) for 
$\bar\varphi^0_{\mu\nu}(t_l;q)$ and (\ref{Deltanres}) for
$\bar\varphi^0(t_r;q)$. This gives us
\begin{eqnarray*}
\bar\varphi^0(t_r\vee t_l;q) &=& \rho(t_l\vee t_r) \varphi^0(\|;q)\\&&
\hspace*{-20mm}
+ \sum_{F(t_r) \Delta t_l} \zeta(t_{l(1)}/t^\circ_{r(1)})
\zeta(t^\bullet_{r(1)}) \bar\varphi^0(t_{l(2)}\vee t_{r(2)};q).
\end{eqnarray*}
The first term $\rho(t_l\vee t_r)$ is consistent with
$\zeta(t_l\vee t_r)$ as defined in Eq.(\ref{zetarho}).
And the other terms are consistent with Eq.(\ref{Deltanres})
using the coproduct Eq.(\ref{Deltarecn}).

Exactly the same substitution leads to Eq.(\ref{Deltafres})
using the coproduct Eq.(\ref{Deltarecf}).

In Ref.\cite{Frabetti}, it will be shown that $\Delta$
is coassociative and defines a Hopf algebra
over the two-coloured planar binary trees.


\section{Conclusion}
The method of Schwinger equations has a number of advantages:
operator-valued distributions are avoided, as
well as indefinite norms and many of the difficult mathematical 
concepts of quantum field theory. As compared to Feynman diagrams,
the method of planar binary trees is more compact and does not
require symmetry factors. Moreover, our treatment of renormalization
does not require a special treatment of the so-called overlapping
divergences. Such a special treatment is even necessary for 
Kreimer's original method of renormalization by rooted trees
\cite{Kreimer}. 

The present work was much inspired by 
Refs.\cite{Kreimer98,Connes,ConnesK,Kreimer} but it was
carried out independently of the
recent and fascinating results by Connes and Kreimer
\cite{CKI,CKII}. Still, our Hopf algebra can
probably be modified to fit into their general framework.
We hope to explore this connection in a forthcoming
publication.

We conclude this paper with a word of caution. We have not actually
proved that each tree is renormalized by the Hopf algebra. 
In other words, we have not proved that all
$\bar\varphi^0(t;q)$ and $\bar\varphi^0_{\lambda\mu}(t;q)$ are finite.
However, we can give an argument that can eventually lead
to a proof of finiteness. 
In the case of quenched QED, there is a single tree ($\Y^n$)
at each order of the perturbation theory. 
Each order of perturbation theory of quenched
QED is renormalized by the standard renormalization of 
Feynman diagrams, so we know that our $\bar\varphi^0(\Y^n;q)$
and $\bar\varphi^0_{\lambda\mu}(t;q)$ are finite, since they
are uniquely fixed by the Schwinger equations.
The other trees are obtained from some $\Y^n$ by insertion
of renormalized photon propagators. Since these insertions
do not introduce new divergences, and particularly no new
overlapping divergence, we can conclude that all trees
are finite. The point that is still not solved is the insertion
of higher-order photon Green functions, such as the 
photon-photon scattering diagrams. Since these diagrams
are finite, they probably do not spoil the convergence,
but we could not prove this generally.

\section{Aknowledgements}
Ch. B warmly thanks Dirk Kreimer for his invitation to Mainz and 
his illuminating introduction to renormalization.
We are very grateful to Dirk Kreimer and David Broadhurst
for their constant help and support.
We thankfully acknowledge the help and encouragement 
from Jean-Louis Loday, Muriel Livernet
and Fr\'ed\'eric Chapoton. Finally, we want to stress that
our understanding of renormalization has greatly benefited from
the lectures by Alain Connes at the Coll\`ege de France.
This is IPGP contribution \#0000.


\section{Appendix 1}
This appendix contains some proofs.

\subsection{Proof of (\ref{uffa}) (the pruning operator is coassociative)}

If $t=\|$ we have $P(\|)=0$, so the identity (\ref{uffa}) holds. 
So, suppose that $t \neq \|$. The reason why the identity (\ref{uffa}) holds 
for $t$ is that applying the recursive definition of $P$, on the successive 
right grafters of $t$, at each step both sides of (\ref{uffa}) coincide 
on the terms which do not involve $P(t')$ for the last right grafter $t'$ 
considered. 
Of course, when we finally meet a right grafter $t'$ such that $P(t')=0$ 
we obtain the equality (\ref{uffa}). 
We develop this idea formally. 

Any tree $t \neq \|$ can be written in a unique way as 
$$
t= t_1 \vee (t_2 \vee (... \vee (t_n \vee \|)...)), 
$$
for some $n \leq |t|+1$. In fact, it suffices to decompose the tree $t$ 
into its left and right grafting trees, then to decompose successively 
the right trees as graftings of two new trees and pick up all their left 
sides, $t_1 := t_l$, $t_2 := (t_r)_l$, $t_3 := ((t_r)_r)_l$ and so on, 
until we meet an undecomposable right side $(...((t_r)_r)...)_r = \|$. 
Since $|t| = |t_1|+|t_2|+...+|t_n| + n-1$ and $|t_i| \geq 0$ for all 
$i=1,...,n$, the procedure must finish for an $n \leq |t|+1$. 

Since $P(t_n \vee \|)=0$, we have 

\begin{eqnarray*}
P(t_{n-1} \vee (t_{n} \vee \|)) 
&=& t_{n-1} \vee \| \otimes t_{n} \vee \|, \\ 
P(t_{n-2} \vee (t_{n-1} \vee (t_{n} \vee \|))) 
&=& t_{n-2} \vee \| \otimes t_{n-1} \vee (t_{n} \vee \|) \\
&+& t_{n-2} \vee (t_{n-1} \vee \|) \otimes t_{n} \vee \| 
\end{eqnarray*}
and
\begin{eqnarray*}
P(t_{n-3} \vee (t_{n-2} \vee (t_{n-1} \vee (t_{n} \vee \|)))) = \hspace{2.5cm} \\
 t_{n-3} \vee \| \otimes t_{n-2} \vee (t_{n-1} \vee (t_{n} \vee \|)) \\
\hspace{2cm} 
+ t_{n-3} \vee (t_{n-2} \vee \|) \otimes t_{n-1} \vee (t_{n} \vee \|) \\
\hspace{2cm} 
+ t_{n-3} \vee (t_{n-2} \vee (t_{n-1} \vee \|)) \otimes t_{n} \vee \|  
\end{eqnarray*}
Thus, for the tree $t= t_1 \vee (t_2 \vee (... \vee (t_n \vee \|)...))$, 
we obtain 
\begin{eqnarray*}
P(t) &=& \sum_{i=1}^{n-1} 
t_1 \vee (t_2 \vee (... \vee (t_i \vee \|)...)) \\
&&\otimes 
t_{i+1} \vee (t_{i+2} \vee (... \vee (t_n \vee \|)...)). 
\end{eqnarray*}
Hence 

\begin{eqnarray*}
&& (P \otimes id) \circ P(t) = \\
&& \qquad = \sum_{j=1}^{n-1}
P(t_1 \vee (t_2 \vee (... \vee (t_i \vee \|)...))) \\
&& \qquad \otimes
t_{i+1} \vee (t_{i+2} \vee (... \vee (t_n \vee \|)...)) \\
&& \qquad = \sum_{j=1}^{n-1} \sum_{i=1}^{j-1}
t_1 \vee (... (t_i \vee \|)...) \\
&& \qquad \otimes
t_{i+1} \vee (... (t_j \vee \|)...) \otimes
t_{j+1} \vee (... (t_n \vee \|)...) \\
&& \qquad = \sum_{1\leq i<j\leq n-1}
t_1 \vee (... (t_i \vee \|)...) \\
&& \qquad \otimes
t_{i+1} \vee (... (t_j \vee \|)...) \otimes
t_{j+1} \vee (... (t_n \vee \|)...) ,
\end{eqnarray*}

and similarly 
\begin{eqnarray*}
&& (id \otimes P) \circ P(t) = \\
&& \qquad = \sum_{k=1}^{n-1}
t_1 \vee (t_2 \vee (... \vee (t_k \vee \|)...)) \\
&& \qquad \otimes
P(t_{k+1} \vee (t_{k+2} \vee (... \vee (t_n \vee \|)...))) \\
&& \qquad = \sum_{k=1}^{n-1} \sum_{l=k+1}^{n-1}
t_1 \vee (... (t_k \vee \|)...) \\
&& \qquad \otimes
t_{k+1} \vee (... (t_l \vee \|)...) \otimes
t_{l+1} \vee (... (t_n \vee \|)...)\\
&&  \qquad = \sum_{1\leq k<l\leq n-1}
t_1 (... \vee (t_k \vee \|)...) \\
&& \qquad \otimes
t_{k+1} \vee (... (t_l \vee \|)...) \otimes
t_{l+1} \vee (... (t_n \vee \|)...) .
\end{eqnarray*}

Hence the identity (\ref{uffa}) holds for any tree.

\subsection{Proof of (\ref{psiphipsi}) and (\ref{Piphi1}) 
  \label{phipsi}}
We use (\ref{phiphi2}) to prove (\ref{psiphipsi}).

\begin{eqnarray*}
D_{\mu\nu}(q)&=& \sum_{t} e_0^{2|t|} \varphi_{\mu\nu}^0 (t;q)
\\&=&
\varphi_{\mu\nu}^0 (\|;q)+\sum_{t_1 t_2} e_0^{2|t_1|+2|t_2|+2} 
\varphi_{\mu\nu}^0 (t_1\vee t_2;q)
\\&=&
\varphi_{\mu\nu}^0 (\|;q)+\sum_{t_1 t_2} e_0^{2|t_1|+2|t_2|+2} 
\varphi^0_{\mu\lambda}(\|\vee t_2;q)
\\&&\times
\varpsi^{0\lambda\lambda'}(\|;q)\varphi^{0}_{\lambda'\nu}(t_1;q)
\\&=&
\varphi_{\mu\nu}^0 (\|;q)+\sum_{t_2} e_0^{2|t_2|+2} 
\varphi^0_{\mu\lambda}(\|\vee t_2;q)
\\&&\times
\varpsi^{0\lambda\lambda'}(\|;q)
D_{\lambda'\nu}(q).
\end{eqnarray*}

We multiply the last equation by $D^{-1}(q)$ on the
right and by ${D^0}^{-1}(q)$ on the left.
This gives us
\begin{eqnarray*}
{[{D^0}^{-1}]}_{\mu\nu}(q) & = & {[D^{-1}]}_{\mu\nu}(q)
+\sum_{t_2} e_0^{2|t_2|+2} {[{D^0}^{-1}]}_{\mu\lambda}(q)\\&&
\varphi^{0\lambda\lambda'}(\|\vee t_2;q)
\varpsi^{0}_{\lambda'\nu}(\|;q).
\end{eqnarray*}
Since $\varphi^{0\lambda\lambda'}(\|\vee t_2;q)$ is
transverse, we can replace 
${[{D^0}^{-1}]}_{\mu\lambda}(q)$ by
$\varpsi^{0}_{\mu\lambda}(\|;q)$ in the above equation.

Using \ref{defDmun}
and the definition of $\Pi_{\mu\nu}(q)$ given in (\ref{defPi})
we obtain
\begin{eqnarray*}
\Pi_{\mu\nu}(q)& = & -\sum_{t_2} e_0^{2|t_2|+2}
\varpsi^{0}_{\mu\lambda}(\|;q)\varphi^{0\lambda\lambda'}(\|\vee t_2;q)
\varpsi^{0}_{\lambda'\nu}(\|;q).
\end{eqnarray*}

We can rewrite this last equation as
\begin{eqnarray*}
\Pi_{\mu\nu}(q)& = &\sum_{|t|>0} e_0^{2|t|}                  
\varpsi^{0}_{\mu\nu}(t;q)= \sum_{t} e_0^{2|t|+2}
\varpsi^{0}_{\mu\nu}(\|\vee t;q),
\end{eqnarray*}
with
$\varpsi^{0\mu\nu}(t_1 \vee t_2;q)=0$ if
$t_1\not= \|$ and 
\begin{eqnarray}
\varpsi^{0}_{\mu\nu}(\|\vee t;q) & = &
-\varpsi^{0}_{\mu\lambda}(\|;q)\varphi^{0\lambda\lambda'}(\|\vee t;q)
\varpsi^{0}_{\lambda'\nu}(\|;q). \label{kedir}
\end{eqnarray}

Finally, we use (\ref{phiblanck}) with $n=0$ and $t_1=\|$
\begin{eqnarray*}
\varphi^0_{\mu\nu}(\|\vee t_2;q)&=&
 -iD^0_{\mu\lambda}(q)\int \frac{\dd^4p}{(2\pi)^4}
  \mathrm{tr}\big[\gamma^\lambda
  \varphi^{1}(t_2;p;\lambda',-q)
  \big]
  \nonumber\\&&\hspace*{0mm}
  \times D^0_{\lambda'\nu}(q).
\end{eqnarray*}
We multiply this equation by 
${D^0}^{-1}(q)$ on the left and on the right,
and we use the fact that $\varphi^0_{\mu\nu}(\|\vee t_2;q)$
is transverse to get, from (\ref{kedir}), the relation
\begin{eqnarray*}
\varpsi^{0\mu\nu}(\|\vee t;q) & = &
i\int \frac{\dd^4p}{(2\pi)^4}
  \mathrm{tr}\big[\gamma^\mu
  \varphi^{1}(t;p;\nu,-q)
  \big].
\end{eqnarray*}

If we sum over trees $t$, the last equation becomes
\begin{eqnarray*}
\Pi^{\lambda\mu}(q)=i e_0^2
  \int \frac{\dd^4p}{(2\pi)^4}
  \mathrm{tr}\big[\gamma^\lambda
  \frac{\delta S(p)}{e_0 \delta A^0_\mu(-q)}\big].
\end{eqnarray*}

\subsection{Proof of (\ref{intpsizero}) \label{deuxg}}
From the definition of self-energy $\varpsi$ (Eq.(\ref{inversepsi})),
we use the definition of the convolution (Eq.(\ref{convolution}))
and of the pruning operator (Eq.(\ref{defP})) to write, for
a tree $t=t_l\vee t_r$
\begin{eqnarray*}
\varpsi^0(t)&=& \varpsi^0(\|)\varphi^0(t)\varpsi^0(\|)
+\varpsi^0(\|)(\varphi^0\star\varpsi^0)(t)\\
&=& \varpsi^0(\|)\varphi^0(t)\varpsi^0(\|)
+\varpsi^0(\|)\varphi^0(t_l\vee \|)\varpsi^0(t_r)\\
&& + \sum_{i=1}^{n(t_r)}
\varpsi^0(\|)\varphi^0(t_l\vee u_i)\varpsi^0(v_i),
\end{eqnarray*}
where $u_i$ and $v_i$ are the trees obtained by
pruning $t_r$ (i.e. $P(t_r)=\sum_i u_i\otimes v_i$).

The last equation will be transformed by using 
the recursive definition of $\varphi^0$ for the trees
$t$, $t_l\vee \|$ and $t_l\vee u_i$.
For $n=0$, Eq.(\ref{phinoirk}) becomes
\begin{eqnarray*}
\varphi^0(t;q)&=&
 i\int \frac{\dd^4 p}{(2\pi)^4} S^0(q)\gamma^\lambda
 \varphi^0_{\lambda\lambda'} (t_l;p)
 \varphi^1(t_r;q-p;\lambda',p).
\end{eqnarray*}

Therefore, we obtain
\begin{eqnarray*}
\varpsi^0(t;q)&=&
-i\int \frac{\dd^4 p}{(2\pi)^4} \gamma^\lambda
 \varphi^0_{\lambda\lambda'} (t_l;p)\\&&\hspace*{-12mm}
\Big[\varphi^1(t_r;q-p;\lambda',p)\varpsi^0(\|;q)+
   \varphi^1(\|;q-p;\lambda',p)\varpsi^0(t_r;q)\\&&+
\sum_{i=1}^{n(t_r)}\varphi^1(u_i;q-p;\lambda',p)
\varpsi^0(v_i;q)\Big].
\end{eqnarray*}
This can also be written
\begin{eqnarray*}
\varpsi^0(t;q)&=&
 i\int \frac{\dd^4 p}{(2\pi)^4} \gamma^\lambda
 \varphi^0_{\lambda\lambda'} (t_l;p)
 g(t_r;q-p;\lambda',p),
\end{eqnarray*}
where
\begin{eqnarray*}
g(t_r;q-p;\lambda',p) &=&
-\varphi^1(t_r;q-p;\lambda',p)\varpsi^0(\|;q)
\nonumber\\&&-
(\varphi^1(q-p;\lambda',p)\star\varpsi^0(q))(t_r)\\&&-
\varphi^1(\|;q-p;\lambda',p)\varpsi^0(t_r;q).
\end{eqnarray*}

It will be useful to transform
$g(t_r;q-p;\lambda',p)$.
To do that, we rewrite the equation for
$\varpsi^1(t;q;\lambda,p)$ given at the end of paragraph
6.4 in Ref.\cite{BrouderEPJC2}, so that it gives
\begin{eqnarray*}
\varphi^0(\|;q)\varpsi^1(t;q;\lambda,p) &=&
\varphi^0(\|;q)\gamma^\lambda\varphi^0(t;q+p)\varpsi^0(\|;q+p)
\\&&\hspace*{-30mm}-
\varphi^0(t;q)\gamma^\lambda-
\varphi^1(t;q;\lambda,p)\varpsi^0(\|;q+p) 
\\&&\hspace*{-30mm}
+\sum_{i=1}^{n(t)}\Big[
\varphi^0(\|;q)\gamma^\lambda \varphi^0(u_i;q+p)\varpsi^0(v_i;q+p)
\\&&\hspace*{-30mm}-
\varphi^0(u_i;q)\varpsi^1(v_i;q;\lambda,p)-
\varphi^1(u_i;q;\lambda,p)\varpsi^0(v_i;q+p)\Big].
\end{eqnarray*}
Now, we replace $\varphi^0(t;q+p)$ by its value
given from Eq.(\ref{inversephi}), we add
$g(t;q;\lambda,p)$ on both sides and we reorder a bit.
\begin{eqnarray*}
g(t;q;\lambda,p) &=&
\varphi^0(\|;q)\varpsi^1(t;q;\lambda,p) +\varphi^0(t;q)\gamma^\lambda
\\&&\hspace*{0mm}+
\varphi^0(\|;q)\gamma^\lambda\varphi^0(t;q+p)\varpsi^0(\|;q+p)
\\&&\hspace*{0mm}-
\varphi^1(\|;q;\lambda,p)\varpsi^0(\|;q+p)
\\&&\hspace*{0mm}
+\sum_{i=1}^{n(t)}
\varphi^0(u_i;q+p)\varpsi^1(v_i;q;\lambda,p).
\end{eqnarray*}
Finally, we note that
$\varphi^1(\|;q;\lambda,p)=\varphi^0(\|;q)\gamma^\lambda\varphi^0(t;q+p)$
and we obtain
\begin{eqnarray*}
g(t;q;\lambda,p) &=&
\varphi^0(\|;q)\varpsi^1(t;q;\lambda,p)+
\varphi^0(t;q)\varpsi^1(\|;q;\lambda,p)\nonumber\\&&
+\sum_{i=1}^{n(t)}
\varphi^0(u_i;q)\varpsi^1(v_i;q;\lambda,p).
\end{eqnarray*}

\section{Appendix2}

In this appendix, we collect the relation between 
bare and renormalized photon and electron $\varphi$ 
up to three loops.

\subsection{Electron Green function for massless QED}
\subsubsection{One loop}
\begin{eqnarray*}
\bar\varphi(\Y)&=&\varphi(\Y)-\zeta_2(\Y)\varphi(\|).
\end{eqnarray*}

\subsubsection{Two loops}
\begin{eqnarray*}
\bar\varphi(\deuxun)&=&\varphi(\deuxun)-\zeta_3(\Y)\varphi(\Y)
  -\zeta_2(\deuxun)\varphi(\|),\\
\bar\varphi(\deuxdeux)&=&\varphi(\deuxdeux)-\zeta_2(\Y)\varphi(\Y)
  -\zeta_2(\deuxdeux)\varphi(\|)+\zeta_2(\Y)^2\varphi(\|).
\end{eqnarray*}

\subsubsection{Three loops}
\begin{eqnarray*}
\bar\varphi(\troisun)&=&\varphi(\troisun)-2\zeta_3(\Y)\varphi(\deuxun)+
  \zeta_3(\Y)^2\varphi(\Y) - \zeta_2(\troisun)\varphi(\|),\\
\bar\varphi(\troisdeux)&=&\varphi(\troisdeux)-\zeta_3(\deuxdeux)\varphi(\Y)-
  \zeta_2(\troisdeux)\varphi(\|),\\
\bar\varphi(\troistrois)&=&\varphi(\troistrois)-\zeta_3(\Y)\varphi(\deuxdeux)-
  \zeta_2(\Y)\varphi(\deuxun)\\&&+\zeta_2(\Y)\zeta_3(\Y)\varphi(\Y)-
   \zeta_2(\troistrois)\varphi(\|)+\zeta_2(\deuxun)\zeta_2(\Y)\varphi(\|),\\
\bar\varphi(\troisquatre)&=&\varphi(\troisquatre)-
  \zeta_3(\Y)\varphi(\deuxdeux)-
  \zeta_2(\deuxun)\varphi(\Y)\\
  &&-\zeta_2(\troisquatre)\varphi(\|)+
  \zeta_2(\deuxun)\zeta_2(\Y)\varphi(\|),\\
\bar\varphi(\troiscinq)&=&\varphi(\troiscinq)-\zeta_2(\Y)\varphi(\deuxdeux)-
   \zeta_2(\deuxdeux)\varphi(\Y)\\&&\hspace*{-6mm}+\zeta_2(\Y)^2\varphi(\Y)+
   [-\zeta_2(\troiscinq)+2\zeta_2(\deuxdeux)\zeta_2(\Y)-
  \zeta_2(\Y)^3]\varphi(\|).
\end{eqnarray*}

\subsection{Photon Green function for massless QED}

\subsubsection{One loop}
\begin{eqnarray*}
\bar\varphi_{\lambda\mu}(\Y)&=&\varphi_{\lambda\mu}(\Y)-
       \zeta_3(\Y)\varphi_{\lambda\mu}(\|).
\end{eqnarray*}

\subsubsection{Two loops}
\begin{eqnarray*}
\bar\varphi_{\lambda\mu}(\deuxun)&=&\varphi_{\lambda\mu}(\deuxun)-
  2\zeta_3(\Y)\varphi_{\lambda\mu}(\Y)
  +\zeta_3(\Y)^2\varphi_{\lambda\mu}(\|),\\
\bar\varphi_{\lambda\mu}(\deuxdeux)&=&\varphi_{\lambda\mu}(\deuxdeux)-
  \zeta_3(\deuxdeux)\varphi_{\lambda\mu}(\|).
\end{eqnarray*}

\subsubsection{Three loops}
\begin{eqnarray*}
\bar\varphi_{\lambda\mu}(\troisun)&=&\varphi_{\lambda\mu}(\troisun)-
  3\zeta_3(\Y)\varphi_{\lambda\mu}(\deuxun)
   \\&&+3\zeta_3(\Y)^2\varphi_{\lambda\mu}(\Y)-
   \zeta_3(\Y)^3\varphi_{\lambda\mu}(\|),\\
\bar\varphi_{\lambda\mu}(\troisdeux)&=&\varphi_{\lambda\mu}(\troisdeux)-
  \zeta_3(\Y)\varphi_{\lambda\mu}(\deuxdeux)-
  \zeta_3(\deuxdeux)\varphi_{\lambda\mu}(\Y)\\&&+ 
  \zeta_3(\deuxdeux)\zeta_3(\Y)\varphi_{\lambda\mu}(\|),\\
\bar\varphi_{\lambda\mu}(\troistrois)&=&\varphi_{\lambda\mu}(\troistrois)-
  \zeta_3(\Y)\varphi_{\lambda\mu}(\deuxdeux)-
  \zeta_3(\deuxdeux)\varphi_{\lambda\mu}(\Y)\\&&+ 
  \zeta_3(\deuxdeux)\zeta_3(\Y)\varphi_{\lambda\mu}(\|),\\
\bar\varphi_{\lambda\mu}(\troisquatre)&=&\varphi_{\lambda\mu}(\troisquatre)-
  \zeta_3(\Y)\varphi_{\lambda\mu}(\deuxdeux)-
  \zeta_3(\troisquatre)\varphi_{\lambda\mu}(\|),\\
\bar\varphi_{\lambda\mu}(\troiscinq)&=&\varphi_{\lambda\mu}(\troiscinq)-
  \zeta_3(\troiscinq)\varphi_{\lambda\mu}(\|).
\end{eqnarray*}

\subsection{Electron self-energy for massless QED}
\subsubsection{One loop}
\begin{eqnarray*}
\bar\varpsi(\Y)&=&\varpsi(\Y)+\zeta_2(\Y)\varpsi(\|).
\end{eqnarray*}

\subsubsection{Two loops}
\begin{eqnarray*}
\bar\varpsi(\deuxun)&=&\varpsi(\deuxun)-\zeta_3(\Y)\varpsi(\Y)
  +\zeta_2(\deuxun)\varpsi(\|),\\
\bar\varpsi(\deuxdeux)&=&\varpsi(\deuxdeux)+\zeta_2(\Y)\varpsi(\Y)
  +\zeta_2(\deuxdeux)\varpsi(\|).
\end{eqnarray*}

\subsubsection{Three loops}
\begin{eqnarray*}
\bar\varpsi(\troisun)&=&\varpsi(\troisun)-2\zeta_3(\Y)\varpsi(\deuxun)+
  \zeta_3(\Y)^2\varpsi(\Y) + \zeta_2(\troisun)\varpsi(\|),\\
\bar\varpsi(\troisdeux)&=&\varpsi(\troisdeux)-\zeta_3(\deuxdeux)\varpsi(\Y)+
  \zeta_2(\troisdeux)\varpsi(\|),\\
\bar\varpsi(\troistrois)&=&\varpsi(\troistrois)-\zeta_3(\Y)\varpsi(\deuxdeux)+
  \zeta_2(\deuxun)\varpsi(\Y)+
   \zeta_2(\troistrois)\varpsi(\|),\\
\bar\varpsi(\troisquatre)&=&\varpsi(\troisquatre)-\zeta_3(\Y)\varpsi(\deuxdeux)+
  \zeta_2(\troisquatre)\varpsi(\|)\\
  &&+\zeta_2(\Y)\varpsi(\deuxun)-\zeta_2(\Y)\zeta_3(\Y)\varpsi(\Y),\\
\bar\varpsi(\troiscinq)&=&\varpsi(\troiscinq)+\zeta_2(\Y)\varpsi(\deuxdeux)+
   \zeta_2(\deuxdeux)\varpsi(\Y)+
   \zeta_2(\troiscinq)\varpsi(\|).
\end{eqnarray*}

\subsection{Vacuum polarization for massless QED}

\subsubsection{One loop}
\begin{eqnarray*}
\bar\varpsi_{\lambda\mu}(\Y)&=&\varpsi_{\lambda\mu}(\Y)-
       \zeta_3(\Y)\varpsi_{\lambda\mu}(\|).
\end{eqnarray*}

\subsubsection{Two loops}
\begin{eqnarray*}
\bar\varpsi_{\lambda\mu}(\deuxun)&=&0,\\
\bar\varpsi_{\lambda\mu}(\deuxdeux)&=&\varpsi_{\lambda\mu}(\deuxdeux)-
  \zeta_3(\deuxdeux)\varpsi_{\lambda\mu}(\|).
\end{eqnarray*}

\subsubsection{Three loops}
\begin{eqnarray*}
\bar\varpsi_{\lambda\mu}(\troisun)&=&0,\\
\bar\varpsi_{\lambda\mu}(\troisdeux)&=&0,\\
\bar\varpsi_{\lambda\mu}(\troistrois)&=&0,\\
\bar\varpsi_{\lambda\mu}(\troisquatre)&=&\varpsi_{\lambda\mu}(\troisquatre)-
  \zeta_3(\Y)\varpsi_{\lambda\mu}(\deuxdeux)-
  \zeta_3(\troisquatre)\varpsi_{\lambda\mu}(\|),\\
\bar\varpsi_{\lambda\mu}(\troiscinq)&=&\varpsi_{\lambda\mu}(\troiscinq)-
  \zeta_3(\troiscinq)\varpsi_{\lambda\mu}(\|).
\end{eqnarray*}

\section{Appendix 3: The first trees}

This appendix gives the Feynman diagrams corresponding
to the first trees for the electron and photon Green
functions.

\begin{fmffile}{des90}

\setlength{\unitlength}{1mm}

\newcommand{\setval}{\fmfset{wiggly_len}{1.5mm}\fmfset{arrow_len}{2.5mm}
\fmfset{arrow_ang}{13}\fmfset{dash_len}{1.5mm}\fmfpen{0.25mm}
\fmfset{dot_size}{0.8thick}}

\newcommand{\scs}{\scriptstyle}
\subsection{Electron Green function}

For the electron Green functions, all electron loops
are oriented anticlockwise and the propagator is oriented
from right to left, as indicated in $\varphi^0(\|)$.
Once the Feynman diagrams for $\varphi^0(t)$ are known,
those for $\varphi^1(t)$ are obtained by summing all
possible insertions of a photon dangling bond at each
electron propagator $ \parbox{10mm}{\begin{center}
  \begin{fmfgraph*}(8,3)
  \setval
  \fmfleft{v1}
  \fmfright{v2}
  \fmf{fermion}{v2,v1}
  \fmfdot{v1,v2}
  \end{fmfgraph*}
  \end{center}}$, and so on for $\varphi^n(t)$.
Notice that the last two diagrams of $\varphi^0(\troisquatre)$
are zero by Furry's theorem. However, they are
useful to generate Feynman diagrams for higher order trees.

\begin{eqnarray*}
\varphi^0(\|) &=& 
  \parbox{10mm}{\begin{center}
  \begin{fmfgraph*}(8,3)
  \setval
  \fmfleft{v1}
  \fmfright{v2}
  \fmf{fermion}{v2,v1}
  \fmfdot{v1,v2}
  \end{fmfgraph*}
  \end{center}} \\
\varphi^0(\Y) &=& 
  \parbox{27mm}{\begin{center}
  \begin{fmfgraph}(22,5)
  \setval
  \fmfforce{0w,0.5h}{v1}
  \fmfforce{1/3w,0.5h}{v2}
  \fmfforce{2/3w,0.5h}{v3}
  \fmfforce{1w,0.5h}{v4}
  \fmf{plain}{v4,v3,v2,v1}
  \fmf{boson,left}{v2,v3}
  \fmfdot{v1,v2,v3,v4}
  \end{fmfgraph}
  \end{center}}
\end{eqnarray*}
\begin{eqnarray*}
\varphi^0(\deuxdeux) &=& 
\parbox{28mm}{\begin{center}
\begin{fmfgraph}(25,5)
\setval
\fmfforce{0w,0.5h}{v1}
\fmfforce{1/5w,0.5h}{v2}
\fmfforce{2/5w,0.5h}{v3}
\fmfforce{3/5w,0.5h}{v4}
\fmfforce{4/5w,0.5h}{v5}
\fmfforce{1w,0.5h}{v6}
\fmf{plain}{v6,v5,v4,v3,v2,v1}
\fmf{boson,left=0.6}{v2,v4}
\fmf{boson,right=0.6}{v3,v5}
\fmfdot{v1,v2,v3,v4,v5,v6}
\end{fmfgraph}
\end{center}}
+
\parbox{28mm}{\begin{center}
\begin{fmfgraph}(25,5)
\setval
\fmfforce{0w,0.5h}{v1}
\fmfforce{1/5w,0.5h}{v2}
\fmfforce{2/5w,0.5h}{v3}
\fmfforce{3/5w,0.5h}{v4}
\fmfforce{4/5w,0.5h}{v5}
\fmfforce{1w,0.5h}{v6}
\fmf{plain}{v6,v5,v4,v3,v2,v1}
\fmf{boson,left=0.6}{v2,v5}
\fmf{boson,left=0.6}{v3,v4}
\fmfdot{v1,v2,v3,v4,v5,v6}
\end{fmfgraph}
\end{center}}
\\&&
+
\parbox{28mm}{\begin{center}
\begin{fmfgraph}(25,5)
\setval
\fmfforce{0w,0.5h}{v1}
\fmfforce{1/5w,0.5h}{v2}
\fmfforce{2/5w,0.5h}{v3}
\fmfforce{3/5w,0.5h}{v4}
\fmfforce{4/5w,0.5h}{v5}
\fmfforce{1w,0.5h}{v6}
\fmf{plain}{v6,v5,v4,v3,v2,v1}
\fmf{boson,left}{v2,v3}
\fmf{boson,left}{v4,v5}
\fmfdot{v1,v2,v3,v4,v5,v6}
\end{fmfgraph}
\end{center}}
\\
\varphi^0(\deuxun) &=& 
\parbox{28mm}{\begin{center}
\begin{fmfgraph}(25,8)
\setval
\fmfforce{0w,0h}{v1}
\fmfforce{0.2w,0h}{v2}
\fmfforce{0.4w,11/16h}{v2b}
\fmfforce{0.8w,0h}{v3}
\fmfforce{0.6w,11/16h}{v3b}
\fmfforce{1w,0h}{v4}
\fmf{plain}{v4,v3,v2,v1}
\fmf{boson,left=0.33}{v2,v2b}
\fmf{boson,right=0.33}{v3,v3b}
\fmf{plain,right}{v2b,v3b,v2b}
\fmfdot{v1,v2,v3,v3b,v2b,v4}
\end{fmfgraph}
\end{center}}
\end{eqnarray*}
\begin{eqnarray*}
\varphi^0(\troisun) &=& 
\parbox{33mm}{\begin{center}
\begin{fmfgraph}(30,8)
\setval
\fmfforce{0w,0h}{v1}
\fmfforce{1/7w,0h}{v2}
\fmfforce{2/7w,11/16h}{v3}
\fmfforce{3/7w,11/16h}{v4}
\fmfforce{4/7w,11/16h}{v5}
\fmfforce{5/7w,11/16h}{v6}
\fmfforce{6/7w,0h}{v7}
\fmfforce{1w,0h}{v8}
\fmf{plain}{v8,v7,v2,v1}
\fmf{plain,right}{v3,v4,v3}
\fmf{plain,right}{v5,v6,v5}
\fmf{boson,left=0.33}{v2,v3}
\fmf{boson}{v4,v5}
\fmf{boson,left=0.33}{v6,v7}
\fmfdot{v1,v2,v3,v4,v5,v6,v7,v8}
\end{fmfgraph}
\end{center}}
\end{eqnarray*}
\begin{eqnarray*}
\varphi^0(\troisdeux) &=& 
\parbox{28mm}{\begin{center}
\begin{fmfgraph}(25,8)
\setval
\fmfforce{0w,0h}{v1}
\fmfforce{0.2w,0h}{v2}
\fmfforce{0.4w,11/16h}{v2b}
\fmfforce{0.5w,1h}{v4b}
\fmfforce{0.5w,6/16h}{v5b}
\fmfforce{0.8w,0h}{v3}
\fmfforce{0.6w,11/16h}{v3b}
\fmfforce{1w,0h}{v4}
\fmf{plain}{v4,v3,v2,v1}
\fmf{boson,left=0.33}{v2,v2b}
\fmf{boson,right=0.33}{v3,v3b}
\fmf{plain,right}{v2b,v3b,v2b}
\fmf{boson}{v4b,v5b}
\fmfdot{v1,v2,v3,v3b,v2b,v4,v4b,v5b}
\end{fmfgraph}
\end{center}}
+
\parbox{28mm}{\begin{center}
\begin{fmfgraph}(25,8)
\setval
\fmfforce{0w,0h}{v1}
\fmfforce{0.2w,0h}{v2}
\fmfforce{0.4w,11/16h}{v2b}
\fmfforce{0.4134w,27/32h}{v4b}
\fmfforce{0.5866w,27/32h}{v5b}
\fmfforce{0.8w,0h}{v3}
\fmfforce{0.6w,11/16h}{v3b}
\fmfforce{1w,0h}{v4}
\fmf{plain}{v4,v3,v2,v1}
\fmf{boson,left=0.33}{v2,v2b}
\fmf{boson,right=0.33}{v3,v3b}
\fmf{plain,right}{v2b,v3b,v2b}
\fmf{boson,right=0}{v4b,v5b}
\fmfdot{v1,v2,v3,v3b,v2b,v4,v4b,v5b}
\end{fmfgraph}
\end{center}}
\\&+&
\parbox{28mm}{\begin{center}
\begin{fmfgraph}(25,8)
\setval
\fmfforce{0w,0h}{v1}
\fmfforce{0.2w,0h}{v2}
\fmfforce{0.4w,11/16h}{v2b}
\fmfforce{0.4134w,17/32h}{v4b}
\fmfforce{0.5866w,17/32h}{v5b}
\fmfforce{0.8w,0h}{v3}
\fmfforce{0.6w,11/16h}{v3b}
\fmfforce{1w,0h}{v4}
\fmf{plain}{v4,v3,v2,v1}
\fmf{boson,left=0.33}{v2,v2b}
\fmf{boson,right=0.33}{v3,v3b}
\fmf{plain,right}{v2b,v3b,v2b}
\fmf{boson,left=0}{v4b,v5b}
\fmfdot{v1,v2,v3,v3b,v2b,v4,v4b,v5b}
\end{fmfgraph}
\end{center}}
\end{eqnarray*}
\begin{eqnarray*}
\varphi^0(\troistrois) &=& 
\parbox{33mm}{\begin{center}
\begin{fmfgraph}(30,8)
\setval
\fmfforce{0w,0.2h}{v1}
\fmfforce{1/7w,0.2h}{v2}
\fmfforce{2/7w,0.8h}{v3}
\fmfforce{3/7w,0.8h}{v4}
\fmfforce{4/7w,0.2h}{v5}
\fmfforce{5/7w,0.2h}{v6}
\fmfforce{6/7w,0.2h}{v7}
\fmfforce{1w,0.2h}{v8}
\fmf{plain}{v8,v7,v6,v5,v2,v1}
\fmf{boson,left}{v6,v7}
\fmf{plain,left}{v3,v4,v3}
\fmf{boson,left=0.33}{v2,v3}
\fmf{boson,left=0.33}{v4,v5}
\fmfdot{v1,v2,v3,v4,v5,v6,v7,v8}
\end{fmfgraph}
\end{center}}
+
\parbox{33mm}{\begin{center}
\begin{fmfgraph}(30,8)
\setval
\fmfforce{0w,0.2h}{v1}
\fmfforce{1/7w,0.2h}{v2}
\fmfforce{5/14w,0.8h}{v3}
\fmfforce{7/14w,0.8h}{v4}
\fmfforce{4/7w,0.2h}{v5}
\fmfforce{5/7w,0.2h}{v6}
\fmfforce{6/7w,0.2h}{v7}
\fmfforce{1w,0.2h}{v8}
\fmf{plain}{v8,v7,v6,v5,v2,v1}
\fmf{boson,right=0.66}{v5,v7}
\fmf{plain,left}{v3,v4,v3}
\fmf{boson,left=0.33}{v2,v3}
\fmf{boson,left=0.33}{v4,v6}
\fmfdot{v1,v2,v3,v4,v5,v6,v7,v8}
\end{fmfgraph}
\end{center}}
\\
&+&
\parbox{28mm}{\begin{center}
\begin{fmfgraph}(25,8)
\setval
\fmfforce{0w,0.2h}{v1}
\fmfforce{0.2w,0.2h}{v2}
\fmfforce{0.4w,0.8h}{v2b}
\fmfforce{0.4w,0.2h}{v2c}
\fmfforce{0.8w,0.2h}{v3}
\fmfforce{0.6w,0.8h}{v3b}
\fmfforce{0.6w,0.2h}{v3c}
\fmfforce{1w,00.2h}{v4}
\fmf{plain}{v4,v3c,v3,v2c,v2,v1}
\fmf{boson,left=0.33}{v2,v2b}
\fmf{boson,right=0.33}{v3,v3b}
\fmf{plain,right}{v2b,v3b,v2b}
\fmf{boson,right}{v2c,v3c}
\fmfdot{v1,v2,v3,v3b,v2b,v4,v2c,v3c}
\end{fmfgraph}
\end{center}}
\end{eqnarray*}
\begin{eqnarray*}
\varphi^0(\troisquatre) &=& 
\parbox{33mm}{\begin{center}
\begin{fmfgraph}(30,8)
\setval
\fmfforce{0w,0.2h}{v1}
\fmfforce{1/7w,0.2h}{v2}
\fmfforce{6/7w,0.2h}{v3}
\fmfforce{2/7w,0.2h}{v4}
\fmfforce{3/7w,0.8h}{v5}
\fmfforce{4/7w,0.8h}{v6}
\fmfforce{5/7w,0.2h}{v7}
\fmfforce{1w,0.2h}{v8}
\fmf{plain}{v8,v7,v4,v3,v2,v1}
\fmf{boson,right=0.3}{v2,v3}
\fmf{plain,left}{v5,v6,v5}
\fmf{boson,left=0.33}{v4,v5}
\fmf{boson,left=0.33}{v6,v7}
\fmfdot{v1,v2,v3,v4,v5,v6,v7,v8}
\end{fmfgraph}
\end{center}}
+
\parbox{33mm}{\begin{center}
\begin{fmfgraph}(30,8)
\setval
\fmfforce{1w,0.2h}{v1}
\fmfforce{6/7w,0.2h}{v2}
\fmfforce{9/14w,0.8h}{v3}
\fmfforce{7/14w,0.8h}{v4}
\fmfforce{3/7w,0.2h}{v5}
\fmfforce{2/7w,0.2h}{v6}
\fmfforce{1/7w,0.2h}{v7}
\fmfforce{0w,0.2h}{v8}
\fmf{plain}{v8,v7,v6,v5,v2,v1}
\fmf{boson,left=0.66}{v5,v7}
\fmf{plain,left}{v3,v4,v3}
\fmf{boson,right=0.33}{v2,v3}
\fmf{boson,right=0.33}{v4,v6}
\fmfdot{v1,v2,v3,v4,v5,v6,v7,v8}
\end{fmfgraph}
\end{center}}
\\&+&
\parbox{33mm}{\begin{center}
\begin{fmfgraph}(30,8)
\setval
\fmfforce{0w,0.2h}{v1}
\fmfforce{1/7w,0.2h}{v2}
\fmfforce{2/7w,0.2h}{v3}
\fmfforce{3/7w,0.2h}{v4}
\fmfforce{4/7w,0.8h}{v5}
\fmfforce{5/7w,0.8h}{v6}
\fmfforce{6/7w,0.2h}{v7}
\fmfforce{1w,0.2h}{v8}
\fmf{plain}{v8,v7,v4,v3,v2,v1}
\fmf{boson,left}{v2,v3}
\fmf{plain,left}{v5,v6,v5}
\fmf{boson,left=0.33}{v4,v5}
\fmf{boson,left=0.33}{v6,v7}
\fmfdot{v1,v2,v3,v4,v5,v6,v7,v8}
\end{fmfgraph}
\end{center}}
+
\parbox{28mm}{\begin{center}
\begin{fmfgraph}(24,8)
\setval
\fmfforce{0w,0.2h}{v1}
\fmfforce{1/6w,0.2h}{v2}
\fmfforce{2/6w,0.2h}{v3}
\fmfforce{11/24w,0.7h}{v4}
\fmfforce{14/24w,43/40h}{vx}
\fmfforce{17/24w,0.7h}{v5}
\fmfforce{5/6w,0.2h}{v6}
\fmfforce{1w,0.2h}{v7}
\fmf{plain}{v7,v6,v3,v2,v1}
\fmf{boson,left=0.33}{v3,v4}
\fmf{boson,left=0.33}{v5,v6}
\fmf{boson,left=0.33}{v2,vx}
\fmf{plain,right}{v4,v5,v4}
\fmfdot{v1,v2,v3,v4,v5,v6,v7,vx}
\end{fmfgraph}
\end{center}}
\\&+&
\parbox{28mm}{\begin{center}
\begin{fmfgraph}(25,8)
\setval
\fmfforce{0w,0.2h}{v1}
\fmfforce{1/6w,0.2h}{v2}
\fmfforce{2/6w,0.2h}{v3}
\fmfforce{11/24w,0.7h}{v4}
\fmfforce{14/24w,13/40h}{vx}
\fmfforce{17/24w,0.7h}{v5}
\fmfforce{5/6w,0.2h}{v6}
\fmfforce{1w,0.2h}{v7}
\fmf{plain}{v7,v6,v3,v2,v1}
\fmf{boson,left=0.33}{v3,v4}
\fmf{boson,left=0.33}{v5,v6}
\fmf{boson,right=0.33}{v2,vx}
\fmf{plain,right}{v4,v5,v4}
\fmfdot{v1,v2,v3,v4,v5,v6,v7,vx}
\end{fmfgraph}
\end{center}}
\end{eqnarray*}
\begin{eqnarray*}
\varphi^0(\troiscinq) &=& 
\parbox{33mm}{\begin{center}
\begin{fmfgraph}(30,5)
\setval
\fmfforce{0w,0.5h}{v1}
\fmfforce{1/7w,0.5h}{v2}
\fmfforce{2/7w,0.5h}{v3}
\fmfforce{3/7w,0.5h}{v4}
\fmfforce{4/7w,0.5h}{v5}
\fmfforce{5/7w,0.5h}{v6}
\fmfforce{6/7w,0.5h}{v7}
\fmfforce{1w,0.5h}{v8}
\fmf{plain}{v8,v7,v6,v5,v4,v3,v2,v1}
\fmf{boson,right=0.6}{v2,v5}
\fmf{boson,left=0.6}{v3,v7}
\fmf{boson,left=0.6}{v4,v6}
\fmfdot{v1,v2,v3,v4,v5,v6,v7,v8}
\end{fmfgraph}
\end{center}}
+
\parbox{33mm}{\begin{center}
\begin{fmfgraph}(30,5)
\setval
\fmfforce{0w,0.5h}{v1}
\fmfforce{1/7w,0.5h}{v2}
\fmfforce{2/7w,0.5h}{v3}
\fmfforce{3/7w,0.5h}{v4}
\fmfforce{4/7w,0.5h}{v5}
\fmfforce{5/7w,0.5h}{v6}
\fmfforce{6/7w,0.5h}{v7}
\fmfforce{1w,0.5h}{v8}
\fmf{plain}{v8,v7,v6,v5,v4,v3,v2,v1}
\fmf{boson,left=0.6}{v2,v6}
\fmf{boson,left}{v4,v5}
\fmf{boson,left=0.6}{v3,v7}
\fmfdot{v1,v2,v3,v4,v5,v6,v7,v8}
\end{fmfgraph}
\end{center}}
\\&+&
\parbox{33mm}{\begin{center}
\begin{fmfgraph}(30,5)
\setval
\fmfforce{0w,0.5h}{v1}
\fmfforce{1/7w,0.5h}{v2}
\fmfforce{2/7w,0.5h}{v3}
\fmfforce{3/7w,0.5h}{v4}
\fmfforce{4/7w,0.5h}{v5}
\fmfforce{5/7w,0.5h}{v6}
\fmfforce{6/7w,0.5h}{v7}
\fmfforce{1w,0.5h}{v8}
\fmf{plain}{v8,v7,v6,v5,v4,v3,v2,v1}
\fmf{boson,left=0.6}{v2,v7}
\fmf{boson,left=0.6}{v3,v5}
\fmf{boson,right=0.6}{v4,v6}
\fmfdot{v1,v2,v3,v4,v5,v6,v7,v8}
\end{fmfgraph}
\end{center}}
+
\parbox{33mm}{\begin{center}
\begin{fmfgraph}(30,5)
\setval
\fmfforce{0w,0.5h}{v1}
\fmfforce{1/7w,0.5h}{v2}
\fmfforce{2/7w,0.5h}{v3}
\fmfforce{3/7w,0.5h}{v4}
\fmfforce{4/7w,0.5h}{v5}
\fmfforce{5/7w,0.5h}{v6}
\fmfforce{6/7w,0.5h}{v7}
\fmfforce{1w,0.5h}{v8}
\fmf{plain}{v8,v7,v6,v5,v4,v3,v2,v1}
\fmf{boson,left=0.6}{v2,v4}
\fmf{boson,right=0.6}{v3,v7}
\fmf{boson,left}{v5,v6}
\fmfdot{v1,v2,v3,v4,v5,v6,v7,v8}
\end{fmfgraph}
\end{center}}
\\&+&
\parbox{33mm}{\begin{center}
\begin{fmfgraph}(30,5)
\setval
\fmfforce{0w,0.5h}{v1}
\fmfforce{1/7w,0.5h}{v2}
\fmfforce{2/7w,0.5h}{v3}
\fmfforce{3/7w,0.5h}{v4}
\fmfforce{4/7w,0.5h}{v5}
\fmfforce{5/7w,0.5h}{v6}
\fmfforce{6/7w,0.5h}{v7}
\fmfforce{1w,0.5h}{v8}
\fmf{plain}{v8,v7,v6,v5,v4,v3,v2,v1}
\fmf{boson,left=0.6}{v2,v7}
\fmf{boson,left}{v3,v4}
\fmf{boson,left}{v5,v6}
\fmfdot{v1,v2,v3,v4,v5,v6,v7,v8}
\end{fmfgraph}
\end{center}}
+
\parbox{33mm}{\begin{center}
\begin{fmfgraph}(30,5)
\setval
\fmfforce{0w,0.5h}{v1}
\fmfforce{1/7w,0.5h}{v2}
\fmfforce{2/7w,0.5h}{v3}
\fmfforce{3/7w,0.5h}{v4}
\fmfforce{4/7w,0.5h}{v5}
\fmfforce{5/7w,0.5h}{v6}
\fmfforce{6/7w,0.5h}{v7}
\fmfforce{1w,0.5h}{v8}
\fmf{plain}{v8,v7,v6,v5,v4,v3,v2,v1}
\fmf{boson,left=0.6}{v2,v5}
\fmf{boson,right=0.6}{v3,v6}
\fmf{boson,left=0.6}{v4,v7}
\fmfdot{v1,v2,v3,v4,v5,v6,v7,v8}
\end{fmfgraph}
\end{center}}
\\&+&
\parbox{33mm}{\begin{center}
\begin{fmfgraph}(30,5)
\setval
\fmfforce{0w,0.5h}{v1}
\fmfforce{1/7w,0.5h}{v2}
\fmfforce{2/7w,0.5h}{v3}
\fmfforce{3/7w,0.5h}{v4}
\fmfforce{4/7w,0.5h}{v5}
\fmfforce{5/7w,0.5h}{v6}
\fmfforce{6/7w,0.5h}{v7}
\fmfforce{1w,0.5h}{v8}
\fmf{plain}{v8,v7,v6,v5,v4,v3,v2,v1}
\fmf{boson,left=0.6}{v2,v4}
\fmf{boson,right=0.6}{v3,v6}
\fmf{boson,left=0.6}{v5,v7}
\fmfdot{v1,v2,v3,v4,v5,v6,v7,v8}
\end{fmfgraph}
\end{center}}
+
\parbox{33mm}{\begin{center}
\begin{fmfgraph}(30,5)
\setval
\fmfforce{0w,0.5h}{v1}
\fmfforce{1/7w,0.5h}{v2}
\fmfforce{2/7w,0.5h}{v3}
\fmfforce{3/7w,0.5h}{v4}
\fmfforce{4/7w,0.5h}{v5}
\fmfforce{5/7w,0.5h}{v6}
\fmfforce{6/7w,0.5h}{v7}
\fmfforce{1w,0.5h}{v8}
\fmf{plain}{v8,v7,v6,v5,v4,v3,v2,v1}
\fmf{boson,left=0.6}{v2,v6}
\fmf{boson,left=0.6}{v3,v5}
\fmf{boson,right=0.6}{v4,v7}
\fmfdot{v1,v2,v3,v4,v5,v6,v7,v8}
\end{fmfgraph}
\end{center}}
\\&+&
\parbox{33mm}{\begin{center}
\begin{fmfgraph}(30,5)
\setval
\fmfforce{0w,0.5h}{v1}
\fmfforce{1/7w,0.5h}{v2}
\fmfforce{2/7w,0.5h}{v3}
\fmfforce{3/7w,0.5h}{v4}
\fmfforce{4/7w,0.5h}{v5}
\fmfforce{5/7w,0.5h}{v6}
\fmfforce{6/7w,0.5h}{v7}
\fmfforce{1w,0.5h}{v8}
\fmf{plain}{v8,v7,v6,v5,v4,v3,v2,v1}
\fmf{boson,right=0.6}{v2,v6}
\fmf{boson,left}{v3,v4}
\fmf{boson,left=0.6}{v5,v7}
\fmfdot{v1,v2,v3,v4,v5,v6,v7,v8}
\end{fmfgraph}
\end{center}}
+
\parbox{33mm}{\begin{center}
\begin{fmfgraph}(30,5)
\setval
\fmfforce{0w,0.5h}{v1}
\fmfforce{1/7w,0.5h}{v2}
\fmfforce{2/7w,0.5h}{v3}
\fmfforce{3/7w,0.5h}{v4}
\fmfforce{4/7w,0.5h}{v5}
\fmfforce{5/7w,0.5h}{v6}
\fmfforce{6/7w,0.5h}{v7}
\fmfforce{1w,0.5h}{v8}
\fmf{plain}{v8,v7,v6,v5,v4,v3,v2,v1}
\fmf{boson,left=0.6}{v2,v7}
\fmf{boson,left=0.6}{v3,v6}
\fmf{boson,left=0.6}{v4,v5}
\fmfdot{v1,v2,v3,v4,v5,v6,v7,v8}
\end{fmfgraph}
\end{center}}
\\&+&
\parbox{33mm}{\begin{center}
\begin{fmfgraph}(30,5)
\setval
\fmfforce{0w,0.5h}{v1}
\fmfforce{1/7w,0.5h}{v2}
\fmfforce{2/7w,0.5h}{v3}
\fmfforce{3/7w,0.5h}{v4}
\fmfforce{4/7w,0.5h}{v5}
\fmfforce{5/7w,0.5h}{v6}
\fmfforce{6/7w,0.5h}{v7}
\fmfforce{1w,0.5h}{v8}
\fmf{plain}{v8,v7,v6,v5,v4,v3,v2,v1}
\fmf{boson,left=0.6}{v2,v4}
\fmf{boson,right=0.6}{v3,v5}
\fmf{boson,left}{v6,v7}
\fmfdot{v1,v2,v3,v4,v5,v6,v7,v8}
\end{fmfgraph}
\end{center}}
+
\parbox{33mm}{\begin{center}
\begin{fmfgraph}(30,5)
\setval
\fmfforce{0w,0.5h}{v1}
\fmfforce{1/7w,0.5h}{v2}
\fmfforce{2/7w,0.5h}{v3}
\fmfforce{3/7w,0.5h}{v4}
\fmfforce{4/7w,0.5h}{v5}
\fmfforce{5/7w,0.5h}{v6}
\fmfforce{6/7w,0.5h}{v7}
\fmfforce{1w,0.5h}{v8}
\fmf{plain}{v8,v7,v6,v5,v4,v3,v2,v1}
\fmf{boson,left}{v2,v3}
\fmf{boson,left=0.6}{v4,v7}
\fmf{boson,left=0.6}{v5,v6}
\fmfdot{v1,v2,v3,v4,v5,v6,v7,v8}
\end{fmfgraph}
\end{center}}
\\&+&
\parbox{33mm}{\begin{center}
\begin{fmfgraph}(30,5)
\setval
\fmfforce{0w,0.5h}{v1}
\fmfforce{1/7w,0.5h}{v2}
\fmfforce{2/7w,0.5h}{v3}
\fmfforce{3/7w,0.5h}{v4}
\fmfforce{4/7w,0.5h}{v5}
\fmfforce{5/7w,0.5h}{v6}
\fmfforce{6/7w,0.5h}{v7}
\fmfforce{1w,0.5h}{v8}
\fmf{plain}{v8,v7,v6,v5,v4,v3,v2,v1}
\fmf{boson,left}{v2,v3}
\fmf{boson,left=0.6}{v4,v6}
\fmf{boson,right=0.6}{v5,v7}
\fmfdot{v1,v2,v3,v4,v5,v6,v7,v8}
\end{fmfgraph}
\end{center}}
+
\parbox{33mm}{\begin{center}
\begin{fmfgraph}(30,5)
\setval
\fmfforce{0w,0.5h}{v1}
\fmfforce{1/7w,0.5h}{v2}
\fmfforce{2/7w,0.5h}{v3}
\fmfforce{3/7w,0.5h}{v4}
\fmfforce{4/7w,0.5h}{v5}
\fmfforce{5/7w,0.5h}{v6}
\fmfforce{6/7w,0.5h}{v7}
\fmfforce{1w,0.5h}{v8}
\fmf{plain}{v8,v7,v6,v5,v4,v3,v2,v1}
\fmf{boson,left=0.6}{v2,v5}
\fmf{boson,left=0.6}{v3,v4}
\fmf{boson,left}{v6,v7}
\fmfdot{v1,v2,v3,v4,v5,v6,v7,v8}
\end{fmfgraph}
\end{center}}
\\&+&
\parbox{33mm}{\begin{center}
\begin{fmfgraph}(30,5)
\setval
\fmfforce{0w,0.5h}{v1}
\fmfforce{1/7w,0.5h}{v2}
\fmfforce{2/7w,0.5h}{v3}
\fmfforce{3/7w,0.5h}{v4}
\fmfforce{4/7w,0.5h}{v5}
\fmfforce{5/7w,0.5h}{v6}
\fmfforce{6/7w,0.5h}{v7}
\fmfforce{1w,0.5h}{v8}
\fmf{plain}{v8,v7,v6,v5,v4,v3,v2,v1}
\fmf{boson,left}{v2,v3}
\fmf{boson,left}{v4,v5}
\fmf{boson,left}{v6,v7}
\fmfdot{v1,v2,v3,v4,v5,v6,v7,v8}
\end{fmfgraph}
\end{center}}
\end{eqnarray*}

\subsection{Photon Green function}

For the photon Green functions, all electron loops
are oriented anticlockwise. This is only indicated
explicitly for $\varphi^0_{\lambda\mu}(\Y)$.

\begin{eqnarray*}
\varphi^0_{\lambda\mu}(\|) &=& 
\parbox{22mm}{\begin{center}
\begin{fmfgraph*}(9,3)
\setval
\fmfleft{v1}
\fmfright{v2}
\fmf{boson}{v1,v2}
\fmfdot{v1,v2}
\end{fmfgraph*}
\end{center}}
\\
\varphi^0_{\lambda\mu}(\Y) &=& 
\parbox{18mm}{\begin{center}
\begin{fmfgraph}(15,5)
\setval
\fmfforce{0w,0.5h}{v1}
\fmfforce{1/3w,0.5h}{v2}
\fmfforce{2/3w,0.5h}{v3}
\fmfforce{1w,0.5h}{v4}
\fmf{boson}{v1,v2}
\fmf{fermion,right}{v2,v3,v2}
\fmf{boson}{v3,v4}
\fmfdot{v1,v2,v3,v4}
\end{fmfgraph}
\end{center}}
\end{eqnarray*}
\begin{eqnarray*}
\varphi^0_{\lambda\mu}(\deuxdeux) &=& 
\parbox{18mm}{\begin{center}
\begin{fmfgraph}(15,5)
\setval
\fmfforce{0w,0.5h}{v1}
\fmfforce{1/3w,0.5h}{v2}
\fmfforce{1/2w,1h}{v2b}
\fmfforce{2/3w,0.5h}{v3}
\fmfforce{1/2w,0h}{v3b}
\fmfforce{1w,0.5h}{v4}
\fmf{boson}{v1,v2}
\fmf{plain,right=0.45}{v2,v3b,v3,v2b,v2}
\fmf{boson}{v3,v4}
\fmf{boson}{v2b,v3b}
\fmfdot{v1,v2,v2b,v3,v3b,v4}
\end{fmfgraph}
\end{center}}+
\parbox{18mm}{\begin{center}
\begin{fmfgraph}(15,5)
\setval
\fmfforce{0w,0.5h}{v1}
\fmfforce{1/3w,0.5h}{v2}
\fmfforce{0.4008w,0.933h}{v2b}
\fmfforce{0.5992w,0.933h}{v2c}
\fmfforce{2/3w,0.5h}{v3}
\fmfforce{1w,0.5h}{v4}
\fmf{boson}{v1,v2}
\fmf{plain,left=0.33}{v2,v2b,v2c,v3}
\fmf{plain,left=1}{v3,v2}
\fmf{boson,right=0.8}{v2b,v2c}
\fmf{boson}{v3,v4}
\fmfdot{v1,v2,v2b,v2c,v3,v4}
\end{fmfgraph}
\end{center}}
\\&&+
\parbox{18mm}{\begin{center}
\begin{fmfgraph}(15,5)
\setval
\fmfforce{0w,0.5h}{v1}
\fmfforce{1/3w,0.5h}{v2}
\fmfforce{2/3w,0.5h}{v3}
\fmfforce{0.4008w,0.067h}{v3b}
\fmfforce{0.5992w,0.067h}{v3c}
\fmfforce{1w,0.5h}{v4}
\fmf{boson}{v1,v2}
\fmf{plain,right=1}{v3,v2}
\fmf{plain,right=0.33}{v2,v3b,v3c,v3}
\fmf{boson,left=0.8}{v3b,v3c}
\fmf{boson}{v3,v4}
\fmfdot{v1,v2,v3b,v3c,v3,v4}
\end{fmfgraph}
\end{center}}
\\
\varphi^0_{\lambda\mu}(\deuxun) &=& 
\parbox{28mm}{\begin{center}
\begin{fmfgraph}(25,5)
\setval
\fmfforce{0w,0.5h}{v1}
\fmfforce{1/5w,0.5h}{v2}
\fmfforce{2/5w,0.5h}{v3}
\fmfforce{3/5w,0.5h}{v4}
\fmfforce{4/5w,0.5h}{v5}
\fmfforce{1w,0.5h}{v6}
\fmf{boson}{v1,v2}
\fmf{plain,right}{v2,v3,v2}
\fmf{boson}{v3,v4}
\fmf{plain,right}{v4,v5,v4}
\fmf{boson}{v5,v6}
\fmfdot{v1,v2,v3,v4,v5,v6}
\end{fmfgraph}
\end{center}}
\end{eqnarray*}
\begin{eqnarray*}
\varphi^0_{\lambda\mu}(\troisun) &=& 
\parbox{33mm}{\begin{center}
\begin{fmfgraph}(30,5)
\setval
\fmfforce{0w,0.5h}{v1}
\fmfforce{1/7w,0.5h}{v2}
\fmfforce{2/7w,0.5h}{v3}
\fmfforce{3/7w,0.5h}{v4}
\fmfforce{4/7w,0.5h}{v5}
\fmfforce{5/7w,0.5h}{v6}
\fmfforce{6/7w,0.5h}{v7}
\fmfforce{1w,0.5h}{v8}
\fmf{boson}{v1,v2}
\fmf{plain,left}{v2,v3,v2}
\fmf{boson}{v3,v4}
\fmf{plain,left}{v4,v5,v4}
\fmf{boson}{v5,v6}
\fmf{plain,left}{v6,v7,v6}
\fmf{boson}{v7,v8}
\fmfdot{v1,v2,v3,v4,v5,v6,v7,v8}
\end{fmfgraph}
\end{center}}
\end{eqnarray*}
\begin{eqnarray*}
\varphi^0_{\lambda\mu}(\troisdeux) &=& 
\parbox{33mm}{\begin{center}
\begin{fmfgraph}(28,8)
\setval
\fmfforce{0w,0.5h}{v1}
\fmfforce{5/28w,0.5h}{v2}
\fmfforce{10/28w,0.5h}{v3}
\fmfforce{15/28w,0.5h}{v4}
\fmfforce{19/28w,1h}{v5}
\fmfforce{19/28w,0h}{v6}
\fmfforce{23/28w,0.5h}{v7}
\fmfforce{1w,0.5h}{v8}
\fmf{boson}{v1,v2}
\fmf{plain,left}{v2,v3,v2}
\fmf{boson}{v3,v4}
\fmf{plain,left}{v4,v7,v4}
\fmf{boson}{v5,v6}
\fmf{boson}{v7,v8}
\fmfdot{v1,v2,v3,v4,v5,v6,v7,v8}
\end{fmfgraph}
\end{center}}
+
\parbox{33mm}{\begin{center}
\begin{fmfgraph}(28,8)
\setval
\fmfforce{0w,0.5h}{v1}
\fmfforce{5/28w,0.5h}{v2}
\fmfforce{10/28w,0.5h}{v3}
\fmfforce{15/28w,0.5h}{v4}
\fmfforce{.55485w,0.75h}{v5}
\fmfforce{.80229w,0.75h}{v6}
\fmfforce{23/28w,0.5h}{v7}
\fmfforce{1w,0.5h}{v8}
\fmf{boson}{v1,v2}
\fmf{plain,left}{v2,v3,v2}
\fmf{boson}{v3,v4}
\fmf{plain,left}{v4,v7,v4}
\fmf{boson}{v5,v6}
\fmf{boson}{v7,v8}
\fmfdot{v1,v2,v3,v4,v5,v6,v7,v8}
\end{fmfgraph}
\end{center}}
\\
&+&
\parbox{33mm}{\begin{center}
\begin{fmfgraph}(28,8)
\setval
\fmfforce{0w,0.5h}{v1}
\fmfforce{5/28w,0.5h}{v2}
\fmfforce{10/28w,0.5h}{v3}
\fmfforce{15/28w,0.5h}{v4}
\fmfforce{.55485w,0.25h}{v5}
\fmfforce{.80229w,0.25h}{v6}
\fmfforce{23/28w,0.5h}{v7}
\fmfforce{1w,0.5h}{v8}
\fmf{boson}{v1,v2}
\fmf{plain,left}{v2,v3,v2}
\fmf{boson}{v3,v4}
\fmf{plain,left}{v4,v7,v4}
\fmf{boson}{v5,v6}
\fmf{boson}{v7,v8}
\fmfdot{v1,v2,v3,v4,v5,v6,v7,v8}
\end{fmfgraph}
\end{center}}
\end{eqnarray*}
\begin{eqnarray*}
\varphi^0_{\lambda\mu}(\troistrois) &=& 
\parbox{33mm}{\begin{center}
\begin{fmfgraph}(28,8)
\setval
\fmfforce{1w,0.5h}{v1}
\fmfforce{23/28w,0.5h}{v2}
\fmfforce{18/28w,0.5h}{v3}
\fmfforce{13/28w,0.5h}{v4}
\fmfforce{9/28w,1h}{v5}
\fmfforce{9/28w,0h}{v6}
\fmfforce{5/28w,0.5h}{v7}
\fmfforce{0w,0.5h}{v8}
\fmf{boson}{v1,v2}
\fmf{plain,left}{v2,v3,v2}
\fmf{boson}{v3,v4}
\fmf{plain,left}{v4,v7,v4}
\fmf{boson}{v5,v6}
\fmf{boson}{v7,v8}
\fmfdot{v1,v2,v3,v4,v5,v6,v7,v8}
\end{fmfgraph}
\end{center}}
+
\parbox{33mm}{\begin{center}
\begin{fmfgraph}(28,8)
\setval
\fmfforce{1w,0.5h}{v1}
\fmfforce{23/28w,0.5h}{v2}
\fmfforce{18/28w,0.5h}{v3}
\fmfforce{13/28w,0.5h}{v4}
\fmfforce{.44515w,0.75h}{v5}
\fmfforce{.19771w,0.75h}{v6}
\fmfforce{5/28w,0.5h}{v7}
\fmfforce{0w,0.5h}{v8}
\fmf{boson}{v1,v2}
\fmf{plain,left}{v2,v3,v2}
\fmf{boson}{v3,v4}
\fmf{plain,left}{v4,v7,v4}
\fmf{boson}{v5,v6}
\fmf{boson}{v7,v8}
\fmfdot{v1,v2,v3,v4,v5,v6,v7,v8}
\end{fmfgraph}
\end{center}}
\\
&+&
\parbox{33mm}{\begin{center}
\begin{fmfgraph}(28,8)
\setval
\fmfforce{1w,0.5h}{v1}
\fmfforce{23/28w,0.5h}{v2}
\fmfforce{18/28w,0.5h}{v3}
\fmfforce{13/28w,0.5h}{v4}
\fmfforce{.44515w,0.25h}{v5}
\fmfforce{.19771w,0.25h}{v6}
\fmfforce{5/28w,0.5h}{v7}
\fmfforce{0w,0.5h}{v8}
\fmf{boson}{v1,v2}
\fmf{plain,left}{v2,v3,v2}
\fmf{boson}{v3,v4}
\fmf{plain,left}{v4,v7,v4}
\fmf{boson}{v5,v6}
\fmf{boson}{v7,v8}
\fmfdot{v1,v2,v3,v4,v5,v6,v7,v8}
\end{fmfgraph}
\end{center}}
\end{eqnarray*}
\begin{eqnarray*}
\varphi^0_{\lambda\mu}(\troisquatre) &=& 
\parbox{33mm}{\begin{center}
\begin{fmfgraph}(28,16)
\setval
\fmfforce{0w,0.5h}{v1}
\fmfforce{6/28w,0.5h}{v2}
\fmfforce{14/28w,0h}{v3}
\fmfforce{14/28w,1/3h}{v4}
\fmfforce{14/28w,2/3h}{v5}
\fmfforce{14/28w,1h}{v6}
\fmfforce{22/28w,0.5h}{v7}
\fmfforce{1w,0.5h}{v8}
\fmf{boson}{v1,v2}
\fmf{plain,left}{v2,v7,v2}
\fmf{boson}{v3,v4}
\fmf{plain,left}{v4,v5,v4}
\fmf{boson}{v5,v6}
\fmf{boson}{v7,v8}
\fmfdot{v1,v2,v3,v4,v5,v6,v7,v8}
\end{fmfgraph}
\end{center}}
+
\parbox{33mm}{\begin{center}
\begin{fmfgraph}(28,16)
\setval
\fmfforce{0w,0.5h}{v1}
\fmfforce{6/28w,0.5h}{v2}
\fmfforce{.2526w,0.75h}{v3}
\fmfforce{.4175w,0.75h}{v4}
\fmfforce{.5825w,0.75h}{v5}
\fmfforce{.7474w,0.75h}{v6}
\fmfforce{22/28w,0.5h}{v7}
\fmfforce{1w,0.5h}{v8}
\fmf{boson}{v1,v2}
\fmf{plain,left}{v2,v7,v2}
\fmf{boson}{v3,v4}
\fmf{plain,left}{v4,v5,v4}
\fmf{boson}{v5,v6}
\fmf{boson}{v7,v8}
\fmfdot{v1,v2,v3,v4,v5,v6,v7,v8}
\end{fmfgraph}
\end{center}}
\\
&+&
\parbox{33mm}{\begin{center}
\begin{fmfgraph}(28,16)
\setval
\fmfforce{0w,0.5h}{v1}
\fmfforce{6/28w,0.5h}{v2}
\fmfforce{.2526w,0.25h}{v3}
\fmfforce{.4175w,0.25h}{v4}
\fmfforce{.5825w,0.25h}{v5}
\fmfforce{.7474w,0.25h}{v6}
\fmfforce{22/28w,0.5h}{v7}
\fmfforce{1w,0.5h}{v8}
\fmf{boson}{v1,v2}
\fmf{plain,left}{v2,v7,v2}
\fmf{boson}{v3,v4}
\fmf{plain,left}{v4,v5,v4}
\fmf{boson}{v5,v6}
\fmf{boson}{v7,v8}
\fmfdot{v1,v2,v3,v4,v5,v6,v7,v8}
\end{fmfgraph}
\end{center}}
\end{eqnarray*}
\begin{eqnarray*}
\varphi^0_{\lambda\mu}(\troiscinq) &=& 
\parbox{21mm}{\begin{center}
\begin{fmfgraph}(20,12)
\setval
\fmfforce{0w,0.5h}{v1}
\fmfforce{4/20w,0.5h}{v2}
\fmfforce{.2573w,.7939h}{v3}
\fmfforce{.4073w,.9755h}{v4}
\fmfforce{.5927w,.9755h}{v5}
\fmfforce{.7427w,.7939h}{v6}
\fmfforce{16/20w,0.5h}{v7}
\fmfforce{1w,0.5h}{v8}
\fmf{boson}{v1,v2}
\fmf{plain,left}{v2,v7,v2}
\fmf{boson,right=0.3}{v3,v5}
\fmf{boson,right=0.3}{v4,v6}
\fmf{boson}{v7,v8}
\fmfdot{v1,v2,v3,v4,v5,v6,v7,v8}
\end{fmfgraph}
\end{center}}
+
\parbox{21mm}{\begin{center}
\begin{fmfgraph}(20,12)
\setval
\fmfforce{0w,0.5h}{v1}
\fmfforce{4/20w,0.5h}{v2}
\fmfforce{.2573w,.7939h}{v3}
\fmfforce{.4073w,.9755h}{v4}
\fmfforce{.5927w,.9755h}{v5}
\fmfforce{.7427w,.7939h}{v6}
\fmfforce{16/20w,0.5h}{v7}
\fmfforce{1w,0.5h}{v8}
\fmf{boson}{v1,v2}
\fmf{plain,left}{v2,v7,v2}
\fmf{boson,right=0.5}{v3,v4}
\fmf{boson,right=0.5}{v5,v6}
\fmf{boson}{v7,v8}
\fmfdot{v1,v2,v3,v4,v5,v6,v7,v8}
\end{fmfgraph}
\end{center}}
+
\parbox{21mm}{\begin{center}
\begin{fmfgraph}(20,12)
\setval
\fmfforce{0w,0.5h}{v1}
\fmfforce{4/20w,0.5h}{v2}
\fmfforce{.2573w,.7939h}{v3}
\fmfforce{.4073w,.9755h}{v4}
\fmfforce{.5927w,.9755h}{v5}
\fmfforce{.7427w,.7939h}{v6}
\fmfforce{16/20w,0.5h}{v7}
\fmfforce{1w,0.5h}{v8}
\fmf{boson}{v1,v2}
\fmf{plain,left}{v2,v7,v2}
\fmf{boson,right=0.5}{v4,v5}
\fmf{boson,right=0}{v3,v6}
\fmf{boson}{v7,v8}
\fmfdot{v1,v2,v3,v4,v5,v6,v7,v8}
\end{fmfgraph}
\end{center}}
\\&+&
\parbox{21mm}{\begin{center}
\begin{fmfgraph}(20,12)
\setval
\fmfforce{0w,0.5h}{v1}
\fmfforce{4/20w,0.5h}{v2}
\fmfforce{.2879w,.8535h}{v3}
\fmfforce{1/2w,1h}{v4}
\fmfforce{.7121w,.8535h}{v5}
\fmfforce{1/2w,0h}{v6}
\fmfforce{16/20w,0.5h}{v7}
\fmfforce{1w,0.5h}{v8}
\fmf{boson}{v1,v2}
\fmf{plain,left}{v2,v7,v2}
\fmf{boson,right=0}{v3,v5}
\fmf{boson,right=0}{v4,v6}
\fmf{boson}{v7,v8}
\fmfdot{v1,v2,v3,v4,v5,v6,v7,v8}
\end{fmfgraph}
\end{center}}
+
\parbox{21mm}{\begin{center}
\begin{fmfgraph}(20,12)
\setval
\fmfforce{0w,0.5h}{v1}
\fmfforce{4/20w,0.5h}{v2}
\fmfforce{.2879w,.8535h}{v3}
\fmfforce{1/2w,1h}{v4}
\fmfforce{.7121w,.8535h}{v5}
\fmfforce{1/2w,0h}{v6}
\fmfforce{16/20w,0.5h}{v7}
\fmfforce{1w,0.5h}{v8}
\fmf{boson}{v1,v2}
\fmf{plain,left}{v2,v7,v2}
\fmf{boson,right=0.3}{v4,v5}
\fmf{boson,right=0}{v3,v6}
\fmf{boson}{v7,v8}
\fmfdot{v1,v2,v3,v4,v5,v6,v7,v8}
\end{fmfgraph}
\end{center}}
+
\parbox{21mm}{\begin{center}
\begin{fmfgraph}(20,12)
\setval
\fmfforce{0w,0.5h}{v1}
\fmfforce{4/20w,0.5h}{v2}
\fmfforce{.2879w,.8535h}{v3}
\fmfforce{1/2w,1h}{v4}
\fmfforce{.7121w,.8535h}{v5}
\fmfforce{1/2w,0h}{v6}
\fmfforce{16/20w,0.5h}{v7}
\fmfforce{1w,0.5h}{v8}
\fmf{boson}{v1,v2}
\fmf{plain,left}{v2,v7,v2}
\fmf{boson,right=0.3}{v3,v4}
\fmf{boson,right=0}{v5,v6}
\fmf{boson}{v7,v8}
\fmfdot{v1,v2,v3,v4,v5,v6,v7,v8}
\end{fmfgraph}
\end{center}}
\\&+&
\parbox{21mm}{\begin{center}
\begin{fmfgraph}(20,12)
\setval
\fmfforce{0w,0.5h}{v1}
\fmfforce{4/20w,0.5h}{v2}
\fmfforce{.2879w,.8536h}{v3}
\fmfforce{.2879w,.1464h}{v4}
\fmfforce{.7121w,.1464h}{v5}
\fmfforce{.7121w,.8536h}{v6}
\fmfforce{16/20w,0.5h}{v7}
\fmfforce{1w,0.5h}{v8}
\fmf{boson}{v1,v2}
\fmf{plain,left}{v2,v7,v2}
\fmf{boson,right=0}{v3,v4}
\fmf{boson,right=0}{v5,v6}
\fmf{boson}{v7,v8}
\fmfdot{v1,v2,v3,v4,v5,v6,v7,v8}
\end{fmfgraph}
\end{center}}
+
\parbox{21mm}{\begin{center}
\begin{fmfgraph}(20,12)
\setval
\fmfforce{0w,0.5h}{v1}
\fmfforce{4/20w,0.5h}{v2}
\fmfforce{.2879w,.8536h}{v3}
\fmfforce{.2879w,.1464h}{v4}
\fmfforce{.7121w,.1464h}{v5}
\fmfforce{.7121w,.8536h}{v6}
\fmfforce{16/20w,0.5h}{v7}
\fmfforce{1w,0.5h}{v8}
\fmf{boson}{v1,v2}
\fmf{plain,left}{v2,v7,v2}
\fmf{boson,right=0}{v3,v5}
\fmf{boson,right=0}{v4,v6}
\fmf{boson}{v7,v8}
\fmfdot{v1,v2,v3,v4,v5,v6,v7,v8}
\end{fmfgraph}
\end{center}}
+
\parbox{21mm}{\begin{center}
\begin{fmfgraph}(20,12)
\setval
\fmfforce{0w,0.5h}{v1}
\fmfforce{4/20w,0.5h}{v2}
\fmfforce{.2879w,.8536h}{v3}
\fmfforce{.2879w,.1464h}{v4}
\fmfforce{.7121w,.1464h}{v5}
\fmfforce{.7121w,.8536h}{v6}
\fmfforce{16/20w,0.5h}{v7}
\fmfforce{1w,0.5h}{v8}
\fmf{boson}{v1,v2}
\fmf{plain,left}{v2,v7,v2}
\fmf{boson,right=0}{v3,v6}
\fmf{boson,right=0}{v4,v5}
\fmf{boson}{v7,v8}
\fmfdot{v1,v2,v3,v4,v5,v6,v7,v8}
\end{fmfgraph}
\end{center}}
\\&+&
\parbox{21mm}{\begin{center}
\begin{fmfgraph}(20,12)
\setval
\fmfforce{0w,0.5h}{v1}
\fmfforce{4/20w,0.5h}{v2}
\fmfforce{.2573w,.2061h}{v3}
\fmfforce{.4073w,.0245h}{v4}
\fmfforce{.5927w,.0245h}{v5}
\fmfforce{.7427w,.2061h}{v6}
\fmfforce{16/20w,0.5h}{v7}
\fmfforce{1w,0.5h}{v8}
\fmf{boson}{v1,v2}
\fmf{plain,left}{v2,v7,v2}
\fmf{boson,left=0.3}{v3,v5}
\fmf{boson,left=0.3}{v4,v6}
\fmf{boson}{v7,v8}
\fmfdot{v1,v2,v3,v4,v5,v6,v7,v8}
\end{fmfgraph}
\end{center}}
+
\parbox{21mm}{\begin{center}
\begin{fmfgraph}(20,12)
\setval
\fmfforce{0w,0.5h}{v1}
\fmfforce{4/20w,0.5h}{v2}
\fmfforce{.2573w,.2061h}{v3}
\fmfforce{.4073w,.0245h}{v4}
\fmfforce{.5927w,.0245h}{v5}
\fmfforce{.7427w,.2061h}{v6}
\fmfforce{16/20w,0.5h}{v7}
\fmfforce{1w,0.5h}{v8}
\fmf{boson}{v1,v2}
\fmf{plain,left}{v2,v7,v2}
\fmf{boson,left=0.5}{v3,v4}
\fmf{boson,left=0.5}{v5,v6}
\fmf{boson}{v7,v8}
\fmfdot{v1,v2,v3,v4,v5,v6,v7,v8}
\end{fmfgraph}
\end{center}}
+
\parbox{21mm}{\begin{center}
\begin{fmfgraph}(20,12)
\setval
\fmfforce{0w,0.5h}{v1}
\fmfforce{4/20w,0.5h}{v2}
\fmfforce{.2573w,.2061h}{v3}
\fmfforce{.4073w,.0245h}{v4}
\fmfforce{.5927w,.0245h}{v5}
\fmfforce{.7427w,.2061h}{v6}
\fmfforce{16/20w,0.5h}{v7}
\fmfforce{1w,0.5h}{v8}
\fmf{boson}{v1,v2}
\fmf{plain,left}{v2,v7,v2}
\fmf{boson,left=0.5}{v4,v5}
\fmf{boson,right=0}{v3,v6}
\fmf{boson}{v7,v8}
\fmfdot{v1,v2,v3,v4,v5,v6,v7,v8}
\end{fmfgraph}
\end{center}}
\\&+&
\parbox{21mm}{\begin{center}
\begin{fmfgraph}(20,12)
\setval
\fmfforce{0w,0.5h}{v1}
\fmfforce{4/20w,0.5h}{v2}
\fmfforce{.2879w,.1465h}{v3}
\fmfforce{1/2w,1h}{v4}
\fmfforce{.7121w,.1465h}{v5}
\fmfforce{1/2w,0h}{v6}
\fmfforce{16/20w,0.5h}{v7}
\fmfforce{1w,0.5h}{v8}
\fmf{boson}{v1,v2}
\fmf{plain,left}{v2,v7,v2}
\fmf{boson,right=0}{v3,v5}
\fmf{boson,right=0}{v4,v6}
\fmf{boson}{v7,v8}
\fmfdot{v1,v2,v3,v4,v5,v6,v7,v8}
\end{fmfgraph}
\end{center}}
+
\parbox{21mm}{\begin{center}
\begin{fmfgraph}(20,12)
\setval
\fmfforce{0w,0.5h}{v1}
\fmfforce{4/20w,0.5h}{v2}
\fmfforce{.2879w,.1465h}{v3}
\fmfforce{1/2w,1h}{v4}
\fmfforce{.7121w,.1465h}{v5}
\fmfforce{1/2w,0h}{v6}
\fmfforce{16/20w,0.5h}{v7}
\fmfforce{1w,0.5h}{v8}
\fmf{boson}{v1,v2}
\fmf{plain,left}{v2,v7,v2}
\fmf{boson,left=0}{v4,v5}
\fmf{boson,left=0.3}{v3,v6}
\fmf{boson}{v7,v8}
\fmfdot{v1,v2,v3,v4,v5,v6,v7,v8}
\end{fmfgraph}
\end{center}}
+
\parbox{21mm}{\begin{center}
\begin{fmfgraph}(20,12)
\setval
\fmfforce{0w,0.5h}{v1}
\fmfforce{4/20w,0.5h}{v2}
\fmfforce{.2879w,.1465h}{v3}
\fmfforce{1/2w,1h}{v4}
\fmfforce{.7121w,.1465h}{v5}
\fmfforce{1/2w,0h}{v6}
\fmfforce{16/20w,0.5h}{v7}
\fmfforce{1w,0.5h}{v8}
\fmf{boson}{v1,v2}
\fmf{plain,left}{v2,v7,v2}
\fmf{boson,left=0}{v3,v4}
\fmf{boson,right=0.3}{v5,v6}
\fmf{boson}{v7,v8}
\fmfdot{v1,v2,v3,v4,v5,v6,v7,v8}
\end{fmfgraph}
\end{center}}
\end{eqnarray*}

\end{fmffile}

\end{document}